\documentclass[referee,a4paper,12pt,traditabstract]{swsc} 

\usepackage{times}
\usepackage[round]{natbib}
\bibpunct{(}{)}{;}{a}{}{,}
\usepackage{enumitem}
\usepackage{graphicx}
\usepackage{hyperref}
\usepackage{eurosym}
\usepackage{wasysym}
\usepackage{verbatim}
\usepackage{fancyhdr}
\usepackage{lineno}
\usepackage[usenames,dvipsnames,svgnames,table]{xcolor}
\usepackage{multirow}
\usepackage{longtable}
\usepackage{afterpage}
\usepackage[para,multiple]{footmisc}
\usepackage{refcount}
\usepackage{rotating}

\hypersetup{colorlinks=true,citecolor=cyan,urlcolor=cyan,linkcolor=blue}

\newcommand{\modif}[1]{{#1}}

\title{A Space Weather Mission Concept: \\
Observatories of the Solar Corona and Active Regions (OSCAR)}

\author{ 
  Antoine Strugarek\inst{1,2} \and 
  Nils Janitzek\inst{3} \and 
  Arrow Lee\inst{4} \and 
  Philipp L\"oschl\inst{5} \and
  Bernhard Seifert\inst{6} \and
  Sanni Hoilijoki\inst{7,8} \and
  Emil Kraaikamp\inst{9} \and
  Alankrita Isha Mrigakshi\inst{3,10} \and
  Thomas Philippe\inst{11} \and
  Sheila Spina\inst{12} \and
  Malte Br\"ose\inst{13} \and
  Sonny Massahi\inst{14} \and
  Liam O'Halloran\inst{15} \and
  Victor Pereira Blanco\inst{16} \and
  Christoffer Stausland\inst{17} \and
  Philippe Escoubet\inst{18} \and 
  G\"unter Kargl\inst{19}}
\institute{
D\'epartement de physique, Universit\'e de Montr\'eal, C.P. 6128
Succ. Centre-Ville, Montr\'eal, Qc, H3C-3J7, Canada \\
\email{\href{mailto:strugarek@astro.umontreal.ca}{strugarek@astro.umontreal.ca}}
\and
Laboratoire AIM Paris-Saclay, CEA/Irfu Universit\'e Paris-Diderot
CNRS/INSU, F-91191 Gif-sur-Yvette, France \and
Institute of Experimental and Applied Physics, University of Kiel,
Kiel, Germany \and
Mullard Space Science Laboratory, University College London, Dorking,
United Kingdom\and
Institut f\"ur Astrophysik, University of Vienna, Vienna, Austria \and
FOTEC - Department of Aerospace Engineering, Wiener Neustadt, Austria\and
Finnish Meteorological Institute, Helsinki, Finland \and
Department of Physics, University of Helsinki, Helsinki, Finland \and
Royal Observatory of Belgium, Brussels, Belgium\and
German Aerospace Centre (DLR), Institute of Aerospace Medicine,
Cologne, Germany \and 
Institut Sup\'erieur de l'A\'eronautique et de l'Espace, Toulouse, France\and
\textit{La Sapienza} - University of Rome, Italy\and
Department of Physics, Free University of Berlin, Berlin, Germany \and
Danish Space Research Institute, Copenhagen, Denmark \and
School of Medicine and Medical Sciences, University College Dublin,
Dublin, Ireland \and 
Dpto. de Astrof\'isica y CC. de la Atm\'osfera, Universidad
Complutense de Madrid, E-28040 Madrid, Spain \and 
Department of Physics, University of Oslo, Norway\and
ESA/ESTEC, Noordwijk, The Netherlands \and
Space Research Institute, Austrian Academy of Sciences, Austria}

\authorrunning{Strugarek, {\emph et al.}}
\titlerunning{Observatories of the Solar Corona and Active Regions (OSCAR)}

\newcommand{\antoine}[1]{{\color{blue}[antoine] #1}}

\newcommand{\philipp}[1]{{\color{cyan}[philipp] #1}}

\definecolor{darkgreen}{RGB}{0,142,128}

\newcommand{\bernhard}[1]{{\color{brown}[bernhard] #1}}

\begin{document}

\abstract{
\noindent

Coronal Mass Ejections (CMEs) and Corotating Interaction Regions (CIRs) are major sources of magnetic storms on Earth and are therefore considered to be the most dangerous space weather events. The Observatories of Solar Corona and Active Regions (OSCAR) mission is designed to identify the 3D structure of coronal loops and to study the trigger mechanisms of CMEs in solar Active Regions (ARs) as well as their evolution and propagation processes in the inner heliosphere. It also aims to provide monitoring and forecasting of geo-effective CMEs and CIRs. OSCAR would contribute to significant advancements in the field of solar physics, improvements of the current CME prediction models, and provide data for reliable space weather forecasting. These objectives are achieved by utilising two spacecraft with identical instrumentation, located at a heliocentric orbital distance of 1~AU from the Sun. The spacecraft will be separated by an angle of 68$^{\circ}$ to provide optimum stereoscopic view of the solar corona. We study the feasibility of such a mission and propose a preliminary design for OSCAR.
}

\maketitle


\section{Introduction}
\label{sec:introduction}

The OSCAR mission concept was conceived during the Alpbach Summer School\footnote{\url{http://www.summerschoolalpbach.at/}} 2013 on space weather over a period of two weeks. We report here the resulting concept that, we believe, is of significant interest for the design of a future space weather oriented space mission. Our mission consists of twin satellites orbiting the Sun at 1~AU with one leading the Earth and the other trailing behind it. They are designed to improve significantly our knowledge of space weather phenomena as well as to develop a space-based space weather forecasting system.

This paper is organised as follows. We first give a quick introduction on the background and motivations that lead to this mission concept (section~\ref{sec:science_background}). We then develop the mission objectives and the associated key scientific requirements of OSCAR (section~\ref{sec:objs_and_reqs}). In section~\ref{sec:instrumentation}, we provide details of the instrumentation needed onboard OSCAR to satisfy those requirements. A preliminary design of the spacecraft for such a payload is given in section~\ref{sec:spacecraft_design}. The mission design, from the orbit selection to the operational modes and the ground segment, is then described in section \ref{sec:mission_design}. In the following section \ref{sec:cost_and_descoping}, an estimate of the mission cost and subsequent descoping options are discussed. We finally conclude our study in section~\ref{sec:conclusions}.\\

\section{Background and Motivation}
\label{sec:science_background}

Space weather describes the changes in the near-Earth ambient plasma that result from solar and cosmic activity. It is a field of major importance in society today as the environmental conditions in the vicinity of Earth severely affect space and ground based systems \citep[see, e.g.,][]{Schwenn:2006gg,Pulkkinen:2007bf}. Most frequent space weather events originate from the Sun. In order to improve our knowledge and anticipation of space weather phenomena one needs to (i) study the origin of those events inside and at the surface of the Sun, (ii) accurately describe their propagation in the interplanetary medium from Sun to Earth and (iii) improve our understanding of how they impact Earth's magnetosphere and atmosphere. 

Solar dynamics lead to the triggering of CMEs and \modif{indirectly to the development of CIRs} that propagate in the solar wind. When they encounter the Earth's magnetosphere, they trigger strong magnetic reconfigurations known as geomagnetic storms. 
CMEs alone cause more than 80\% of geomagnetic storms and thus represent a severe threat for modern technology \citep{Zhang:2007ki,Liu2014:aa}. 
CMEs originate from the complex magnetic structures within the solar ARs. These regions are created by magnetic flux tubes arising from beneath the photosphere, ultimately forming loops in the corona that are anchored to the solar photosphere \citep[for a review, see][]{Fan:2009kw}.
Magnetic shears and stresses within the photosphere lead to an increase of the energy stored in the coronal loops, often triggering magnetic reconnection inside them and leading to explosive events called \textit{flares} \modif{which emit} energetic particles and radiate in the X-ray and extreme ultra-violet (EUV) bands. In some particular configurations, CMEs are triggered and erupt from the AR \citep[see, e.g.,][]{Nitta:2013iv}. \modif{Note that CMEs are not systematically correlated with flares \citep{Webb:2012er}, but the majority of large flares are observed to be followed by CMEs \citep{Yashiro:2005bx}.} Multiple models have been developed to pinpoint the exact triggering mechanism(s) of CMEs, but their unification is still a major challenge in solar astrophysics \citep[see][and references therein]{Zuccarello:2013ia}. \modif{Such unification would greatly help space weather forecasting programs to anticipate CME launches from the solar surface, and discriminate the potential threats further in advance. Hence, the understanding of the trigger of CMEs is a major goal for the progress of solar physics as well as the development of space weather forecasting systems.}

Once triggered, the CMEs propagate outwards from the Sun in the corona and can reach a distance of 1~AU in a time ranging from 14~hours up to 5~days \citep{Chen:2011gq}. The fast moving CMEs, when geo-directed, are generally considered to be the most dangerous space weather events and are also the most difficult to anticipate due to their high propagation velocity. In addition, CMEs can either accelerate or decelerate during their propagation from Sun to Earth \citep{Gopalswamy:2000ef} \modif{depending on the ambient solar wind in which they evolve and on their initial energy \citep{Manoharan:2006it}}. Hence, forecasts have to use either well-cadenced observations of the heliosphere or solar wind models that accurately take the acceleration/deceleration processes into account. The best well-cadenced heliospheric data today comes from the Solar Terrestrial Relations Observatory (STEREO) mission which proved to be very useful for the prediction of CME arrival time at Earth. The ability to combine direct images of the heliosphere from the Sun to 1~AU from different points of view with the STEREO spacecraft lead to significant novel capabilities in the context of space weather forecasting. \modif{To cite a few, \citet{Mostl:2014iv} showed that the use of heliospheric images were extremely effective in reconstructing and forecasting high-speed solar wind streams. Using the stereoscopic capabilities, \citet{Davies:2013iv} recently developed a robust technique to determine time profiles and propagation direction of transients in the solar wind. These observations were even shown to provide the capability to determine the excitation site of solar energetic particle observed at 1~AU and to relate this site to a particular CME event \citep[see, \textit{e.g.,}][]{Rouillard:2012ft}.} However, because of the choice of orbits for the two STEREO satellites, the data best suited for stereoscopic observations is available for a very limited amount of time. Hence no reliable long-term forecasting system of CMEs can be set up from those observations.

CIRs are produced when the fast solar wind catches up with the slow wind. They consist of high pressure regions that co-rotate with the Sun in the solar wind. They eventually lead to the formation of co-rotating shocks in the supersonic wind, generally at distances larger than 2~AU \citep{Richardson:2004cx}. Depending on their magnetic field orientation with respect to the Earth's magnetic field, CIRs can transfer part of their energy to the magnetosphere which, in turn, causes weak to moderate magnetic storms. Some of these storms can cause significant damage, not only to space technology but also to communication, transportation, and electrical power systems. 

There are currently missions like STEREO and the Solar Dynamics Observatory (SDO) which aim to study -- at least some aspects of -- CMEs and CIRs. While these missions are still providing important new insights for our understanding of the solar corona, none of them truly tackle the hard task of providing a space-based, reliable and long-term space weather forecasting system. In addition, STEREO's stereographic images demonstrated the large gain we obtain from observing the Sun with simultaneous different points of view \citep[for recent reviews, see][and references therein]{Bemporad:2009gf,Zuccarello:2013ia}. Hereby we propose a new mission concept, OSCAR, which aims to  provide new and decisive stereoscopic data that will allow us to finally identify how CMEs are triggered \modif{and how to forecast them}. With our mission design, for the first time \modif{and during its whole lifetime, an efficient forecast of geo-effective CMEs and CIRs will be operating in near-real time. Taking advantage of the position of the spacecraft, CMEs will be monitored and forecasted with remote sensing instruments and CIRs will be forecasted thanks to in-situ measurements before reaching the Earth.}

\section[Science Objectives and Requirements]{Combining Science
  Objectives with Near-Real Time Forecasts}
\label{sec:objs_and_reqs}

The OSCAR mission addresses the difficult challenge of space weather forecasting, from the initiation of interplanetary CMEs (hereafter generically refered to as CMEs) to their coupling with Earth's magnetosphere. In this context, we are interested in the most energetic CMEs and CIRs that can affect the terrestrial environment and human life. In this section we define the mission objectives (section \ref{sec:mission_obj}) as well as the associated scientific key requirements (section \ref{sec:science_req}).

\subsection{Mission Objectives}
\label{sec:mission_obj}

We break down the mission objectives into one primary objective and two secondary objectives. 

\begin{itemize}
\itemsep0em

\item[1]\modif{\textbf{Unveil the trigger mechanism(s) of CMEs in active
      regions and their robust forecasting indices 
}
}

The primary objective is to \modif{provide data to efficiently
  forecast the onset of CMEs in active regions on the solar
  surface. This has to be achieved through the} study of the 3D
structure of coronal loops
\modif{to unveil the} physical trigger mechanism(s) of CMEs. 
The highly energetic CMEs that we are interested in are strongly
associated with M- and X-class flares \citep[only 10\% of the X-class flares
are not associated with CMEs, see][]{Yashiro:2005bx}. 
We will observe the magnetic field of sunspots, the 3D structure
of coronal loops and the flaring process in ARs at the onset of
CMEs. We know that flares and their associated CMEs (when these are
present) occur on a time scale from minutes to hours
\citep{Shibata:2011kd}. High cadenced observations \modif{--one of the
major challenges for this mission--} of these three quantities 
will allow us to sample the whole eruption process \modif{with unprecedented}
details. This will in turn provide strong constraints on the various
physical models of solar eruptions \citep{Zuccarello:2013ia}
\modif{and allow the identification of key plasma quantities as robust
trackers of the trigger of CMEs.}
\modif{In addition,} we will observe hundreds of ARs during the
lifetime of OSCAR and \modif{be able to} create
a catalogue of AR topologies, associated with their ability to trigger
CMEs. \modif{This catalogue will be used to validate theoretical models
  and numerical simulations in terms of necessary conditions
  --\textit{e.g.}, in terms of magnetic helicity in the magnetic
  structure of the AR or temperature profiles across the flux tubes--
  for the trigger of a CME. Hence, 
it will be highly valuable to further guide the future forecasting
of CME triggers on the Sun.}\\

\item[2.a] \textbf{Provide the necessary data for a near-real time
    forecasting geo-affecting/effective CMEs and CIRs}

  Real-time estimates of arrival time for geo-effective CMEs remain
  rather inaccurate today \citep{Davis:2011cn} and strongly depend on solar
  wind models. The aim of OSCAR is to provide data for accurate
  prediction of arrival time of those CMEs \citep[either with a data-driven
  or a purely empirical model, e.g.][]{Howard:2008hj}. 

  CIRs produce 13\% of geomagnetic storms \citep{Zhang:2007ki}. Because they can	
  last a long time and even repeat themselves after a 27-day solar
  rotation, they represent a threat for space based infrastructure
  \citep{Borovsky:2006ec}. In-situ measurements are consequently required to observe them. Since they co-rotate with the Sun, their propagation speed requires distant in-situ observation to be able to forecast them. With OSCAR, we intend to provide data for
  reliable forecasts of CIRs. 

\modif{Altogether the OSCAR mission will provide 6-hour updated
  forecasting data of ocurring CMEs based on high-cadenced
  remote-sensing measurements and of CIRs based on well-located
  in-situ measurements. This will ensure a forecasting window of
  about 8 hours for the fastest CMEs and of 2 full days for CIRs. Thus,
  OSCAR will sustain for all its life-time a reliable forecasting
  for CMEs and CIRs.} 

\item[2.b] \textbf{Enhance our understanding of spatial structures of
    CMEs and CIRs at 1 AU}
  
  Our knowledge of the composition, geometry and magnetic field  of geo-effective CIRs and CMEs is mainly based on local measurements in the vicinity of Earth (except for the STEREO measurements).
With the OSCAR mission we will be able to cover an angle of 68$^{\circ}$ for CME in-situ measurements at 1~AU. In addition OSCAR will be able to
trace the evolution of CIRs on a short timescale
of 5.5 days, which was shown by STEREO to be the shortest typical
timescale for major changes in the solar wind structure \citep{GomezHerrero:2011eo}.

Combining this large set of in-situ measurements with the
remote-sensing techniques described above, OSCAR shall be able to
enhance our basic understanding of space weather relevant CME/CIR
aspects such as (i) CME propagation out to 1~AU
\citep{DalLago:2013jx}, (ii) CME shock acceleration of energetic
electrons \citep{Simnett:2002ig}, (iii) interaction of CMEs and CIRs
at 1~AU \citep{GomezHerrero:2011eo} and (iv) CIR shock acceleration of
low energetic ions, which were found to be an additional precursor for
geomagnetic storms \modif{\citep[although in general the shock structure of CIRs 
fully develops much beyond 1~AU, it was shown by][that some
inner CIRs can be fully developed and are at the origin
of some large geomagnetic storms]{Smith:2004kx}}.

\end{itemize}

Finally, OSCAR also provides numerous targets of opportunity for
studying space weather events from the lower corona to 1~AU with
multiple line of sights observations and multiple in-situ
measurements. We choose to focus here on the main objectives of OSCAR
\modif{which are directly related to the forecast of space
  weather events (see list above)}
to clearly outline the feasibility of this mission concept. We
leave a more exhaustive objectives list \modif{--with more
  science-oriented aspects--} for further studies.

\begin{figure}[ht]
\centering
\includegraphics[width=0.7\linewidth]{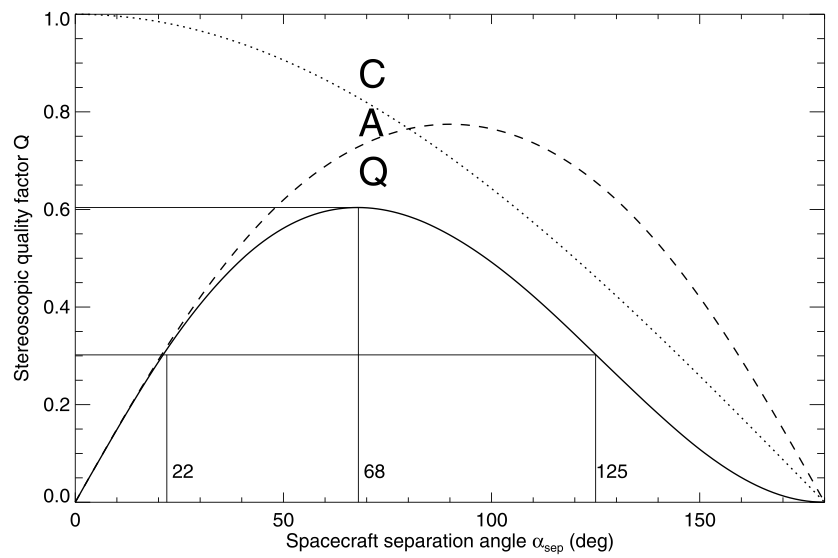}
\caption{Quality factor Q of stereoscopic triangulation as a function
  of the spacecraft separation angle $\alpha_{sep}$, which is a
  function of the accuracy A of triangulated stereoscopic positions
  and the stereoscopic correspondence quality factor C. The best
  quality (within a factor of 2) occurs in the range of $\alpha_{sep}
  =$ 22$^\circ$-125$^\circ$ with an optimal separation angle of
  68$^\circ $. \modif{Figure re-used from \cite{Aschwanden:2012jc} with the
  authorization from Springer.}\label{fig:angles_3D_reconstruction}}
\end{figure}

\subsection{Scientific Requirements}
\label{sec:science_req}

In order to fulfill our objectives, we detail the
scientific requirements for each of them in table~\ref{tab:requirements}. We separate (top level) global requirements
from (second level) quantitative requirements. We give
hereafter short justifications for some key requirements, which will be used to identify the instruments required for OSCAR (section \ref{sec:instrumentation}).

The stereographic observations (required for any 3D reconstruction of coronal loop structures) are one of the cornerstones of our
mission. Thanks to the experience gained from STEREO, we can
precisely define an optimal separation angle between the two lines of
view for the three-dimensional reconstruction of coronal loops in an
AR. \citet{Aschwanden:2012jc} demonstrated that the quality of a stereoscopic
reconstruction depends on both (i) the quality of
correspondance between the two images and (ii) the accuracy of
the triangulation process. The former decreases as the separation
angle decreases while the latter is maximized for an angle of
90$^\circ$. The product of the two factors gives the overall
quality factor of the reconstruction that is given in
figure \ref{fig:angles_3D_reconstruction}. An optimal angle is found
around 68$^{\circ}$ and an acceptable stereoscopic reconstruction is
obtained at a separation angle between 22$^{\circ}$ and
125$^{\circ}$. The onset of a CME occurs on a time scale of tens of
seconds in a coronal structure. Hence, our ambitious primary objective
requires high cadenced 
-- high resolution 
(see table \ref{tab:requirements}) observations of coronal loops (and
associated photospheric magnetic fields) in ARs. 

The major requirements for our secondary objectives are (i)
the ability to provide sufficient regular data for an efficient
forecast and (ii) to accurately measure the propagation and expansion of CMEs (from the lower corona to 1~AU) and the evolution of the spatial structure of CIRs (at 1~AU) with a combination of remote sensing and in-situ observations.
The former requires relatively high-cadenced data (images of the Sun-Earth
heliosphere every 2~hours) to ensure that the fastest CMEs that reach Earth within approximately 15~hours can be accurately forecasted.
We refer the reader to table \ref{tab:requirements} (and references therein) for more details regarding the other specific requirements. Some requirements apply to several objectives, in which case they have not been repeated for sake of simplicity.

\begin{longtable}{|p{3cm}|p{6cm}|p{6cm}|}
\caption{Scientific requirements for the OSCAR mission\label{tab:requirements}}\\
\hline
  \textbf{Objective} & \textbf{Top level requirements}    &
  \textbf{Second level requirements} 
  \\ \hline\hline
  \endfirsthead
  \hline
  \textbf{Objective} & \textbf{Top level requirements}    &
  \textbf{Second level requirements} 
  \\ \hline\hline
  \endhead
\multirow{1}{3cm}{1 - \modif{Trigger mechanism(s) of CMEs in AR and their forecasting indices}
  } & 
  Stereographic
  view of coronal loops at different heights in the lower corona  & The
  separation angle shall be between 22 and 125$^{\circ}$, as
  close to 68 degrees as possible (figure \ref{fig:angles_3D_reconstruction})\\ \cline{2-3}
  & Capture the time scale of flares and of the triggering sequence
  of strong CMEs & Time resolution of 5 seconds for coronal
  loop images \\ \cline{2-3}
  & Resolve distinct coronal loops in ARs & Spatial resolution
  better than 500~km in the solar upper transition region \\ \cline{2-3}
  & Synchronized stereographic images to ensure
  a proper 3D reconstruction during the eruption process &
  The two spacecraft shall be synchronised with a precision of
  0.1~s \\  \cline{2-3}
  & Observe the photospheric vector magnetic field in ARs
  & Spatial
  resolution better than 750~km, with a precision of 0.1~G for
  the longitudinal field and 20~G for the transverse field  \\ \cline{2-3}
  & The duration of OSCAR shall ensure high statistics for the CME triggering & The duration of the mission shall be no less than
  5~years \\

\hline \hline

\multirow{1}{3cm}{2.a - Provide data for the forecast of geo-affecting CMEs} & 
	Track geo-directed CMEs over the whole Sun-Earth distance &
Observations from the lower corona to 1~AU  shall be possible	\\ 
\cline{2-3}
  &  Determine the shape, direction, and velocity of the leading
  edge of the CME (common with 2.b) & The cadence shall be of 2 hours and stereoscopic view shall be available\footnote{\label{ft:h08hj}\citet{Howard:2008hj}}   
\\ \cline{2-3}
  &  Sufficient data to forecast the arrival time of all geo-directed CMEs &
  The data shall enable a 2~days forecast updated every
  6~hours
  \\ 

\hline \hline

\multirow{1}{3cm}{2.a - Provide data for the forecast of geo-effective CIRs} & 
  In-situ measurements of geo-affecting CIRs before they reach  Earth &
  One spacecraft shall be positioned on a keplerian orbit close to 1~AU following Earth 
  \\ \cline{2-3} & 
  	Guarantee sufficient warning time 	
  &
Minimal separation between the Earth-following spacecraft and Earth of 29.7$^\circ$ (warning time of 2.25 days)
  \\ \cline{2-3}
  &  In-situ measurements of the magnetic field (common with 2.b)
 & The vector magnetic field shall be measured within the range $\pm$200~nT with an accuracy of 0.1~nT and an operational time resolution of 1~minute\footnote{\citet{Zwickl:1998fu}\label{ft:Z98}}\\ \cline{2-3}
  & Measurement of the in-situ solar wind proton plasma parameters (speed, temperature, density; common with 2.b) &The solar wind protons shall be measured up to a speed of 1000~km/s (with a 5\% accuracy) with an operational time resolution of 1~minute \footnotemark[\getrefnumber{ft:Z98}]  
  \\

\hline\hline

\multirow{1}{3cm}{2.b Measure the propagation and spatial expansion of CMEs out to 1~AU} &
  Measure continuously the plasma parameters and the magnetic field 
  at 1~AU (in CMEs and CIRs)
  & The solar wind proton speed, 
  temperature and density, the electron 3D velocity distribution, prominent ion charge states of C,O,Si,Fe, and magnetic field shall be measured every 15 minutes\footnote{\citet{Gosling:1992cq,Gloeckler:1998fa,McComas:1998fy,Webb:2012er,DalLago:2013jx}} 
  \\ \cline{2-3}
& Measure high energetic electrons accelerated at CME shock fronts close to the sun
&	The
	energetic electrons passing the spacecraft at 1~AU in the range 40~keV--300~keV shall shall be measured every minute\footnote{\label{ft:Go98}\citet{Gold:1998fc}}\footnote{\citet{Simnett:2002ig,Kahler:2005ky}} 
    \\

\hline \hline

\multirow{1}{3cm}{2.b Measure the evolution of CIR spatial structures at 1~AU} 
& Remote-sensing and tracking of CIRs within 
 one Carrington rotation & CIRs shall be monitored in the heliosphere over an elongation angle of 150$^{\circ}$ over 20 days
 (brightness sensitivity $<3\,\times 10^{-16}$ the solar brightness)\footnote{\citet{Howard:2008fx,Tappin:2009br}} 
  \\ \cline{2-3}
& Multi-point observations at 1~AU to detect changes in the 
structure of CIRs within one Carrington rotation & Minimum seperation angle of
66$^{\circ}$ in longitude to catch major
longitudinal changes, latitudinal separation $<$5$^{\circ}$ \footnote{\label{ft:M09GH11}\citet{Mason:2009cg,GomezHerrero:2011eo}}
  \\ \cline{2-3}
& Measure low energetic ion events accelerated at CIR shock fronts at or beyond 1~AU & The protons and alpha particles in the energy range 50~keV--4~MeV shall be measured every minute\footnotemark[\getrefnumber{ft:Go98}]$^,$\footnotemark[\getrefnumber{ft:M09GH11}]
\\ \hline
\end{longtable}

\begin{figure}[htbp]
\centering
\includegraphics[width=0.7\linewidth]{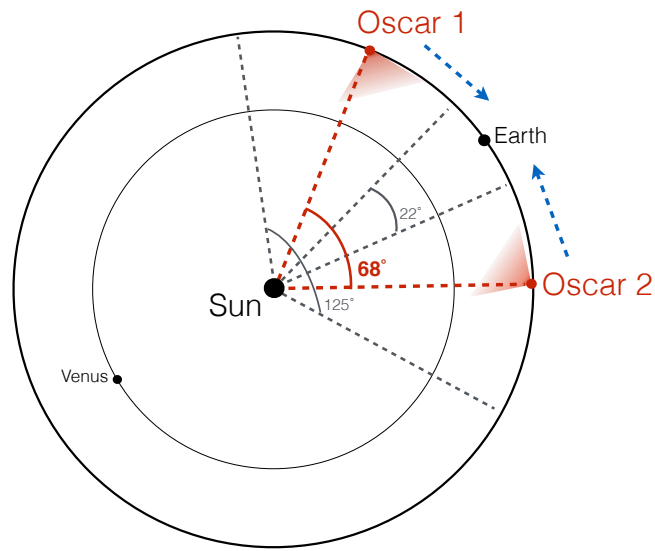}
\caption{Schematic of the OSCAR mission orbital configuration as viewed from the top of the ecliptic plane. The twin spacecraft orbit the Sun at 1~AU with Oscar 1 leading the Earth and Oscar 2 trailing behind it. They are separated by an angle of 68$^\circ$ to cover stereoscopically the photosphere, the upper transition region, the corona and the heliosphere to 1~AU (see section \ref{sec:mission_design} for more details).\label{fig:schematic}}
\end{figure}

Anticipating the detailed design we give in
section \ref{sec:mission_design}, the combination of second level requirements
for objectives (1) and (2a) suggests the use of two identical
spacecraft separated by an angle of 68$^{\circ}$ orbiting the Sun in the Earth orbit. Hence, one can easily imagine one spacecraft leading the Earth and the other trailing behind it with a separation angle of 34$^{\circ}$. A schematic of this design is
given in figure \ref{fig:schematic}. This design also fulfills
all the other scientific requirements listed in table \ref{tab:requirements}. As an example, the satellite orbiting behind  Earth will measure CIRs 2.25 days before they hit Earth. This design also happens to be
achievable with moderate technological developments, as will be made clear in sections \ref{sec:spacecraft_design}-\ref{sec:mission_design}.

\section{Instrumentation}
\label{sec:instrumentation}

Each spacecraft will carry an identical \modif{set} of instruments for the purpose of investigating the Sun and the heliosphere. A suite of telescopes will image the space between the solar surface and 1~AU on the Sun-Earth line almost continuously. Another suite of sensors will measure the in-situ particle and magnetic field environment at 1~AU. An overview of the planned instruments, with estimates of their mass and power consumption, is given in table \ref{tab:instruments}. In order to demonstrate the feasibility of OSCAR, we particulary outline the heritage of our planned instruments. Should this mission concept be further explored, we give paths of improvements for the critical instruments of OSCAR.

\begin{sidewaystable}[htbp]   
  \begin{center}
    \caption{OSCAR instrumentation \label{tab:instruments}}
    \renewcommand{\arraystretch}{1.2}
    \begin{tabular}[h]{|c|c|c|c|c|c|c|c|}
      \hline
 \begin{tabular}[c]{@{}c@{}}\textbf{Instrument} \\ \textbf{(Heritage)}\end{tabular}
 & \textbf{Measurement} & \textbf{Ranges} & \textbf{Resolution} & 
 \begin{tabular}[c]{@{}c@{}}\textbf{Sampling} \\ \textbf{rate}\end{tabular} 
 & \begin{tabular}[c]{@{}c@{}}\textbf{Necessary} \\ \textbf{development}\end{tabular}
 & \begin{tabular}[c]{@{}c@{}}\textbf{Mass} \\ \textbf{[kg]}\end{tabular}
 & \begin{tabular}[c]{@{}c@{}}\textbf{Power} \\ \textbf{[W]}\end{tabular}
 \\
     \hline\hline
      2xMAG &  Interplanetary & $\pm$200~nT & 0.1~nT & 1/16~s & No changes & 1.9$^\mathrm{a}$ & 1.5$^\mathrm{a}$ \\
      (SolO MAG) & B-field vector&  &  & & & \modif{(per unit)} & \modif{(per unit)}\\
     \hline
     SWPM & SW protons, alphas, & $0.26$--36~keV/q  & 5\% ($\Delta E/E$) & 1~min & \modif{Minor changes} & \modif{11.3}$^\mathrm{b,c}$ & \modif{8.7}$^\mathrm{b,c}$\\
     (\modif{STEREO SWEA}  & SW electrons & \modif{1~eV--3~keV} & \modif{17\% ($\Delta E/E$)} & \modif{2~s} & & & \\
 ACE SWEPAM + & SW C,O,Si,Fe ions & 0.49--100~keV/nuc& 6.5\% ($\Delta E/E$) & 12~min & & & \\
 ACE SWICS) &  & & & & & & \\
      \hline
EPM & low energy protons, & 46~keV--4.8~MeV & 8~ch. logscale:   & 1~min & Minor changes & 12.8$^\mathrm{b}$ & 3.8$^\mathrm{b}$\\
(ACE EPAM)       &  alpha particles&  & 11~keV--2.8~MeV  & & & & \\
 & high energy  & 40~keV--350~keV & 4~ch. logscale:  & 1 min & & & \\
      & electrons &  & 17~keV--175~keV & & & & \\
\hline 
      PIM & B-field vector &FOV: Full solar disk & 2048$\times$2048~pix & 45--60~s & Major changes & 29.9$^{*,\mathrm{d}}$ & 28.3$^{*,\mathrm{d}}$ \\
     (SolO PHI)  & at photosphere  & $\lambda$: 617.3~nm& 720~km & & & & \\
\hline
      EUVARI & EUV images &FOV: Full solar disk & 4096$\times$4096~pix & 5~s & New development & 23.5$^{*,\mathrm{d}}$  & 18.0$^{*,\mathrm{d}}$ \\
      (New) &  & $\lambda$: 9.4, 17.1, 21.1~nm & 500~km & & & & \\
            &  & $T$: 6, 0.6, 2 MK & & & & & \\ 
\hline
COR & Thompson  &FOV: 2--15~R$_{\astrosun}$ & 1024$\times$1024 & 15~min & No changes & 11.0$^\mathrm{e}$ & 15.0$^\mathrm{e,f}$ \\
 (STEREO COR 2)& scattered light & $\lambda$: visible light & 15~arcsec/pix & & & & \\
\hline
HI 1+2 & Thompson  & FOV: 12--\modif{277}~R$_{\astrosun}$ & 1024$\times$1024 & 1~h~/~2~h & No changes & 15.0$^{\mathrm{e},\mathrm{f}}$ & 15.0$^{\mathrm{e},\mathrm{f}}$ \\
       (STEREO HI 1+2)& scattered light & $\lambda$: visible light & 70-240~arcsec/pix & & & &\\
     \hline 
\multicolumn{8}{l}{* Estimates from similar instruments}
\end{tabular}
\end{center}
\begin{list}{}{}
\item[] $^\mathrm{a}$ \citet{Carr:2007tj} $\;$ $^\mathrm{b}$ \citet{Stone:1998ea} $\;$ $^\mathrm{c}$ \citet{Sauvaud:2008aa} $\;$ $^\mathrm{d}$  \href{http://sci.esa.int/jump.cfm?oid=48985}{Solar Orbiter Red book} 
$\;$ $^\mathrm{e}$ \citet{Howard:2008fx} \\$\;$ $^\mathrm{f}$ Hancock, private communications
\end{list}
\end{sidewaystable}

\subsection{Remote Sensing Instrumentation}
 
The package of remote-sensing instruments will consist of 4 instruments carrying in total 6 telescopes. Together they will be able to cover the photosphere, the upper transition region, the corona and the heliosphere to beyond 1~AU.


\paragraph{\textbf{Photospheric Imager (PIM)} --}
The Photospheric Imager will provide 2D-maps of the magnetic field vector in the photosphere by measurements of the Zeeman-effect. Similar instruments like SDO's HMI have been built in the past, whereas the upcoming Solar Orbiter's Polarimetric and Heliospheric Imager (PHI) will serve as heritage \citep{Gandorfer:2011dg} in this case. Unlike PHI, only one telescope is required for PIM since the PHI's High Resolution Telescope's (HRT) of 200~km at 0.28~AU already fulfills the spatial resolution as well as the cadence requirements when constantly positioned at a 1~AU orbit. PHI's restricted field of view of 16.8$'$ has to be doubled to observe the full solar disk  at 1~AU, which has to be resolved in further development.
With this setup PIM will be able to observe the spectral line of neutral iron at 617.3~nm with a spatial resolution of about 720~km in the photosphere and a cadence of 45-60~s, which are both within the requirements presented earlier in table \ref{tab:requirements}.
The soon to be flight proven PHI instrument onboard Solar Orbiter will provide a good base for developing the PIM instrument. The changes of design would mostly be focused on the single telescope adaption to the new orbit and the extension of the field of view. A higher observation cadence, although not mandatory, could be very valuable to our understanding of erupting ARs.

\paragraph{\textbf{EUV Active Region Imager (EUVARI)} --}
The EUV imager will provide simultaneous measurements of the full solar disk in two different wavelengths on each spacecraft. Since an instrument with these properties and suitable mass does not exist, technological developments will be needed to produce an instrument that fulfills our requirements. This new instrument will consist of two telescopes. 
One of these will observe the lower corona in the 17.1~nm wavelength, and the other will be able to switch between 9.4 and 21.1~nm using a filter wheel. These wavelength bands cover the footprints of the loops and the loops themselves. They can also be used to detect solar flares (see section \ref{sec:op_mode_ground_seg} for an operational use of those filters). 
To provide more flexibility, to simplify the design and provide some redundancy, both telescopes will be equipped with filter wheels covering each of the three wavelengths.

Significant heritage is available for such an instrument: the SDO, STEREO and Solar Orbiter missions have all flown, or will fly, instruments with some of the characteristics matching our requirements. For instance, the EUV imager on Solar Orbiter will have two telescopes with a combined mass of 23.5~kg, but the filters and fields of view must be changed to suit an orbit at 1~AU.  However, the development and testing of this instrument will be the greatest design challenge of the mission (see section \ref{sec:crit_tech_req} for details). The simultaneous stereoscopic observations -- at high resolution -- of the photospheric magnetic field and the upper coronal structure in ARs will be the key to characterize in details the trigger of CMEs. 

\paragraph{\textbf{Coronagraph (COR)} --} 
The coronagraph (COR) is an externally occulted Lyot coronagraph, which will be capable of imaging the electron density of the solar corona through observing the polarised brightness of Thomson scattered light. To reduce instrument development costs, we can rely on the COR2 instrument flying on the STEREO spacecraft. COR will provide data on early CME propagation between 2 and 15~R$_{\astrosun}$ for first hazard assessments. The instrument is designed to observe moderately fast CMEs moving with speeds of about 750~km/s, which covers the average initial CME speeds as seen in LASCO data \citep{Manoharan:2011hh}. \modif{Like its heritage instrument, it will produce a three--polarisation image sequence in visible light. This will enable to observe the polarised brightness of the coronal plasma, which can be directly related to the distribution of electrons \citep[using, \textit{e.g.}, the solar rotational tomography method, see][]{Frazin:2005aa}.}
To observe the average CME event, short exposure times of $<$4~s are necessary to avoid pixel smear. Faster events can also be resolved by binning pixels \citep{Howard:2008fx}.

\begin{figure}
\centering
\includegraphics[width=1\linewidth]{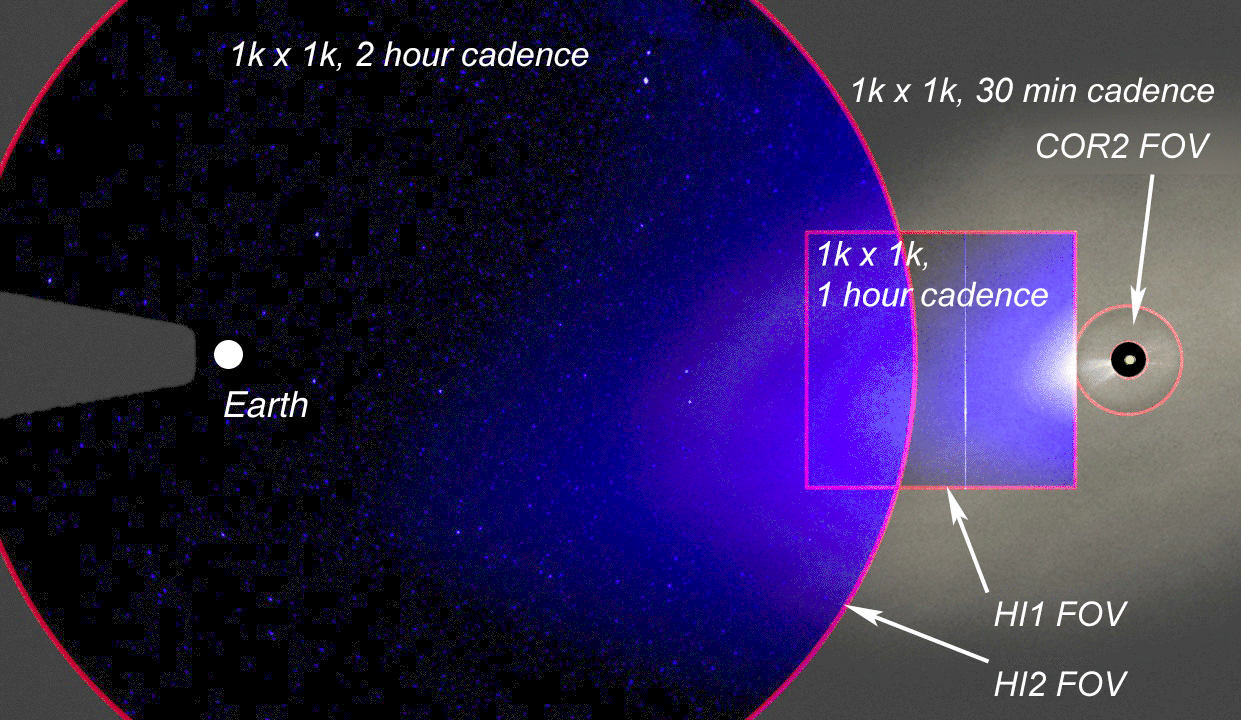}
\caption{The combined field of view of the OSCAR coronal and heliospheric imagers, adapted from the \href{http://stereo.gsfc.nasa.gov/gallery/item.php?id=sciencevisuals&iid=5}{STEREO gallery}. The Earth, not to scale, is labelled for reference.\label{fig:fovs}}
\end{figure}

\paragraph{\textbf{Heliospheric Imagers (HI)} --}
Given the success of the stereoscopic reconstruction with STEREO's Heliospheric Imager (HI) \citep[see, e.g.,][]{Lugaz:2012gz}, the same instrument is also planned to be reused aboard the OSCAR spacecraft. The HI will take over CME observation once they are out of the coronagraph's field of view, following CMEs from 12 to \modif{277}~R$_{\astrosun}$, corresponding to a 85$^{\circ}$ field of view \modif{on the Sun-Earth line}. It consists of the two cameras HI-1 and HI-2 to cover the decreasing intensity of visible Thomson scattered light during the CME's propagation. The opening angles also correspond to typical average CME widths which have been found to range between 47$^{\circ}$ and 61$^{\circ}$ from solar minimum to maximum \citep{Yashiro:2004fo}. Due to the different intensities, the exposure times vary between HI-1 and HI-2. Exposure times of 12--20~s with typically 150 exposures per image for HI-1 and 60--90~s exposure time with tyically 100 exposures per image \modif{allow} a nominal image cadence of 60 and 120~min for this instrument \citep{Howard:2008fx}. Using the COR and HI the volume in which CMEs can be viewed then reaches from 2~R$_{\astrosun}$ to beyond 1~AU, as seen in figure \ref{fig:fovs}.

\subsection{In-Situ Instrumentation}

The in-situ instrumentation aboard each of the OSCAR spacecraft consists of two
identical magnetometers and two different particle instruments, one measuring the solar wind particles, the other one \modif{the shock-accelerated high energy electrons, low energy protons and
alpha particles}. The overall mass of the in-situ instrumentation package is estimated to be 29.4~kg, the power consumption will be 17.0~W.

\paragraph{\textbf{Solar Wind Particle Monitor (SWPM)} --}

The SWPM  will measure the 1D velocity distribution functions (VDFs) of
solar wind protons, alpha-particles and the more abundant charge states of 
certain heavier elements (C, O, Si, Fe) in the solar wind as well as the 3D-VDF of solar wind electrons. Due to the large differences between the expected particle fluxes, the proton/alpha measurement and the heavy ion measurement will be done separately with two different sensors SWPM-AP and SWPM-HI both mounted on the spacecraft body looking towards the Sun. 
\modif{SWPM-AP will measure protons and alphas in the velocity range between 180~km/s
and 2100~km/s. Thus the velocity range is extended compared to regular solar wind speeds in order to measure not only CIRs but even high CME bulk speeds.This design is directly taken from its heritage instrument ACE/SWEPAM  which has an energy-per-nucleon range between 260~eV/nuc and 36~keV/nuc and a relative energy-per-nucleon resolution of 5\% \citep{McComas:1998fy}. SWPM-HI will be able to measure the charge states of carbon, oxygen, silicon and iron with up to 100~kev/nuc in order to determine locally the stream interfaces in CIRs and transient CMEs. The third sensor SWPM-E finally will measure the solar wind electrons. As heritage instruments for SWPM-E we propose STEREO/SWEA \citep{Sauvaud:2008aa}, which can determine 3D velocity distributions of electrons in the energy range between 1~eV and 3~keV.  As its heritage instrument SWPM-E will be mounted on the spacecraft boom.}

\paragraph{\textbf{Energetic Particle Monitor (EPM)} --}
The EPM instrument consists of two identical sensor pairs which will both measure low energetic proton and alpha-particles covering ion energies from 46~keV up to 4.8~MeV, and electron energies between 40~keV and 350~keV. One sensor pair \textit{EPM-1} will be tilted (-45$^{\circ}$) in longitude with respect to the sun-spacecraft line to capture
the high energetic electrons that are expected to propagate along the spiraling interplanetary magnetic field. 
The second sensor pair EPM-2 will point 180$^{\circ}$ away from the first sensor.
This design, implying a rough spatial resolution within the ecliptic, will enable us to (i) identify anisotropic electron events, which are 
the relevant ones in the context of CME shock accelerated particles \citep{Simnett:2002ig} and (ii) distinguish between 
low energetic ions accelerated inside or beyond 1~AU. 
The heritage instrument is  ACE/EPAM \citep{Gold:1998fc}, but note that the sensor mounting would have to be adapted to the 3-axis-stabilized OSCAR spacecraft as described above.

\paragraph{\textbf{Magnetometers (MAG)} --} 
The MAG instrument consists of two identical fluxgate magnetometers. Each of them  will measure all three components of the local interplanetary magnetic field vector in the energy range $\pm$200~nT with an absolute accuracy of 0.1~nT and an operational time resolution of 1 minute for space weather forecasting. Subsecond time resolved data will
be available for scientific studies of the magnetic field, such as wave and turbulence phenomena in the solar wind. The magnetometers will be synchronized and both mounted on the spacecraft boom in a distance of 1.0~m and 2.25~m from the spacecraft body, respectively. This will allow us to reconstruct the magnetic background of the spacecraft itself. The Solar Orbiter MAG \citep{Carr:2007tj} will serve as heritage instrument for the MAG instrument. The combination of MAG and SWPM data will provide the necessary data for a reliable CIRs forecasts in real-time.

\subsection{Critical Technology Requirements}
\label{sec:crit_tech_req}

\modif{
Technically a large proportion of the risk relies in the successful
implementation of the EUV Active Region Imager. To minimise it the
EUVARI telescope will be based on three already developed EUV imagers,
which by the estimated time of launch will each be
flight-proven. Since the data produced by this instrument are critical
for mission success, redundancy in certain crucial
wavelengths imaging will be built into the design in case of damage
during launch or flight.}

\modif{Another large proportion of the risk to our mission lies in the
requirement to successfully implement major changes in the PIM
instrument compared to its heritage instrument PHI: doubling the
16.8$'$ field of view of PHI --to ensure a full disk field of view while
maintaining the high spatial and time resolution-- is the critical
challenge for the successful realisation of this instrument. However,
abandoning the two-telescopes design of PHI should save weight and make
room for PIM specific changes, therefore keeping it compact and
lightweight.  
}

\section{Spacecraft Design}
\label{sec:spacecraft_design}

\begin{figure*}[!htbp]
\centering
\includegraphics[width=\linewidth]{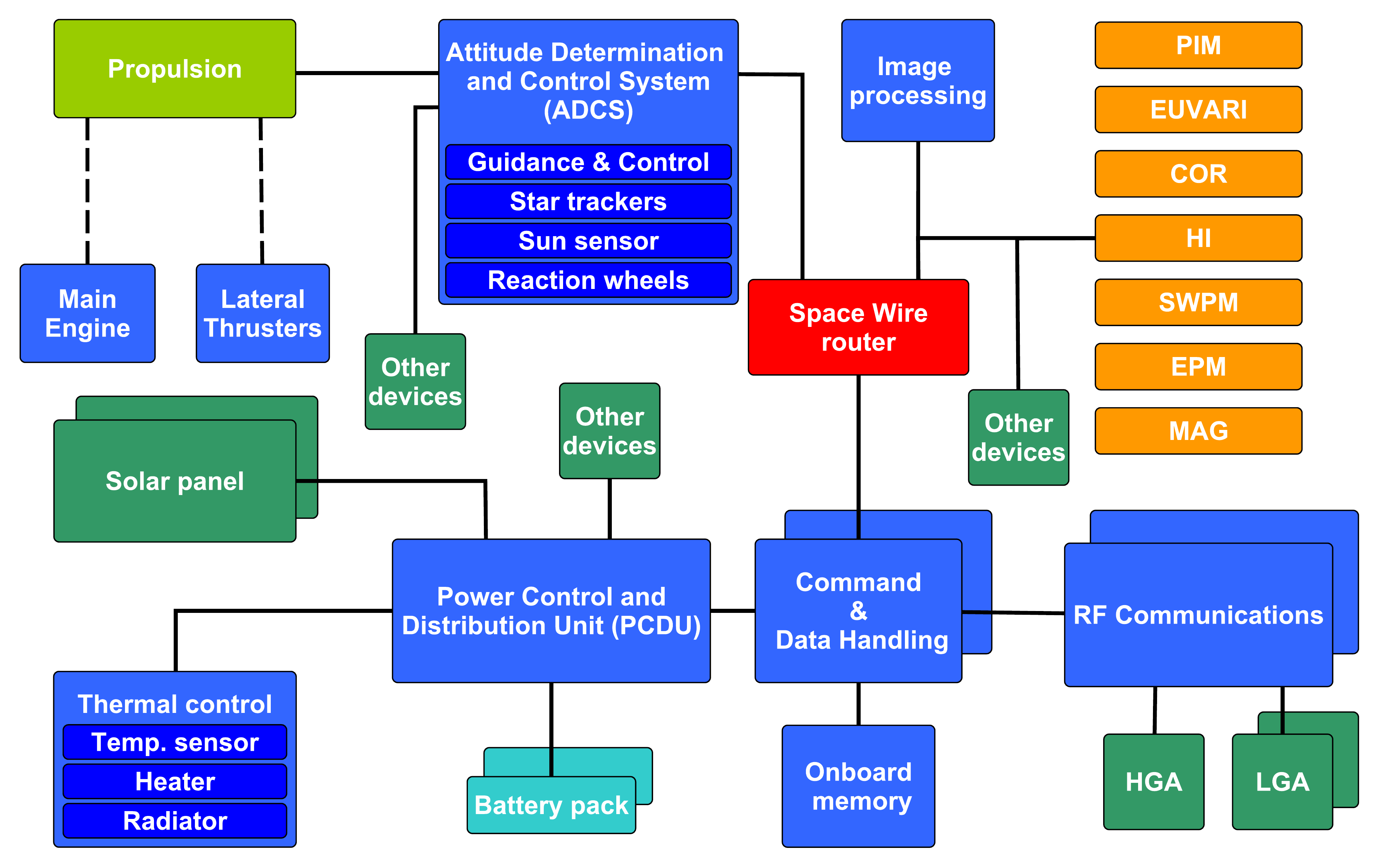}
\includegraphics[width=\linewidth]{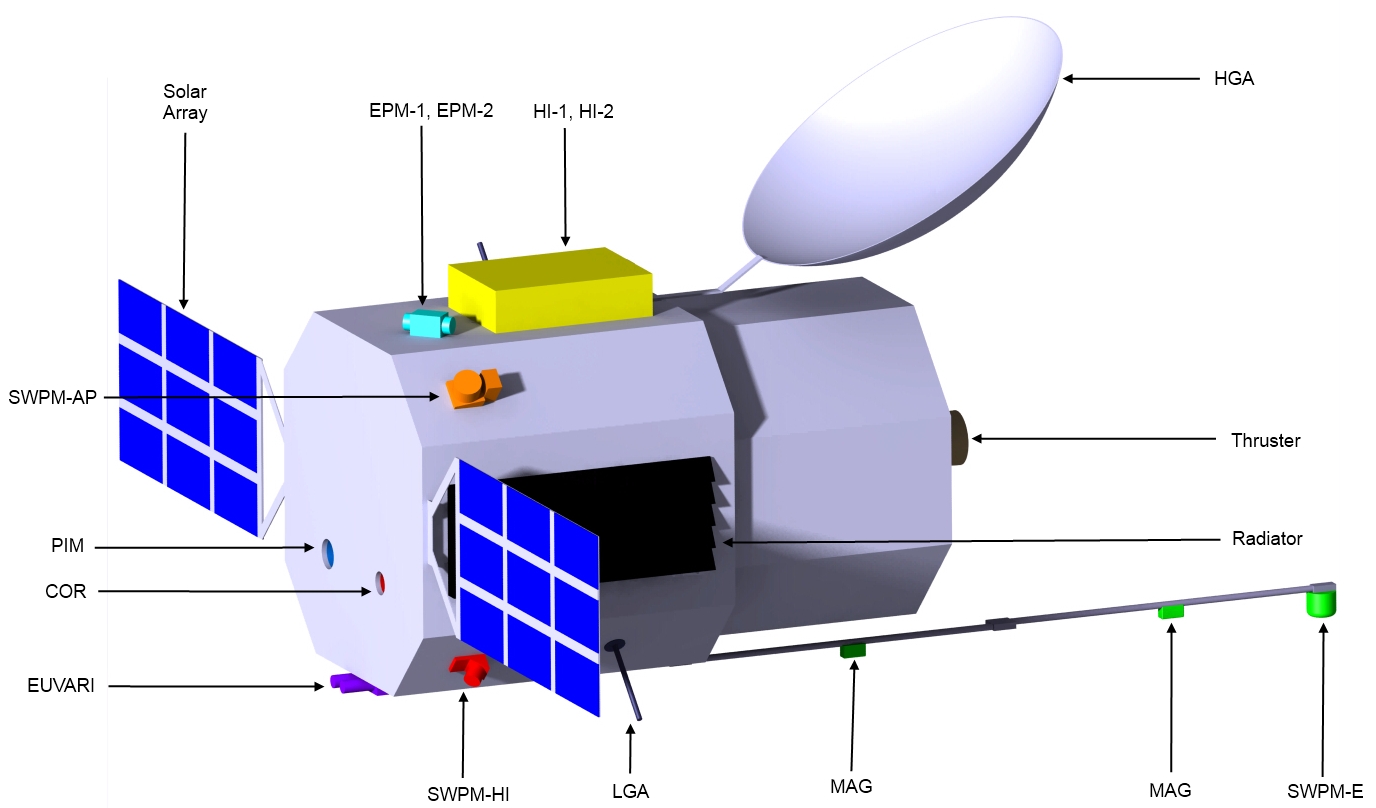}
\caption{System architecture and spacecraft design. The estimated dimensions of the (folded) spacecraft are 1.6 $\times$ 1.6 $\times$ 2.7~m. \label{fig:power_system}}
\end{figure*}

We give now a brief tour of the design of the OSCAR twin spacecraft. The system architecture and satellite design are shown in figure \ref{fig:power_system}. The OSCAR spacecraft are 3-axis stabilised spacecraft that actively point the imaging instruments towards the Sun. The Attitude Determination and Control System (ADCS) computes the attitude and if necessary utilises reaction wheels and lateral thrusters to alter the orientation. Solar panels are utilised to harvest energy which is processed towards the Power Control and Distribution Units (PCDUs) and stored in the batteries. The communication subsystem consists of two redundant X-band transceivers connected to a high gain antenna and two low gain antennas.

\subsection{Mass Budget}

Table \ref{tab:mass} summarises the mass budget for each OSCAR spacecraft. The estimated mass for each spacecraft is \modif{580.5}~kg including -- on top of sub-system margins -- an additional margin of 20\%. If one Soyuz launcher is used for both spacecraft (see section \ref{sec:orbit_and_phases}), an unused mass of \modif{28.2}~kg for each spacecraft provides us a comfortable margin.

For structure, thermal control, onboard computer (OBC) and data handling (DH) a margin of 10\% is provisioned. A margin of 5\% is assumed for the subsystems ADCS and EPS thanks to space heritage. The margins of scientific payload and telemetry, tracking and command (TT\&C) result from the single margins of the susbystems components, based on the standard ESA margin policy. The weight of harness is estimated to 5\% of the net spacecraft mass.

\begin{table}
\centering
\caption{Dry mass budget. The margins are averages based on the standard ESA margin policy.\label{tab:mass}}
\begin{tabular}{|l|c|c|c|}
\hline
\textbf{Subsystem}  & 
\begin{tabular}[c]{@{}c@{}}\textbf{Nominal Mass} \\ \textbf{[kg]}\end{tabular} 
& 
\begin{tabular}[c]{@{}c@{}}\textbf{Margin} \\ \textbf{[\%]}\end{tabular} 
 & 
 \begin{tabular}[c]{@{}c@{}}\textbf{Sum} \\ \textbf{[kg]}\end{tabular} 
 \\
\hline
\hline
Structure  & 149.0 &		10.0 &		163.9 \\
\hline
Payload$^*$ & \modif{110.5} &		10.1 &		\modif{121.6} \\
\hline
TT\&C  &	23.1 &	5.6 &		24.4\\
\hline
ADCS  &	32.1 &	5.0 &	33.7 \\
\hline
OBC\&DH &	25.0 &	5.0 &		27.5 	\\
\hline
EPS	&	18.1 &	5.0 &	19.0 \\
\hline
Thermal &	15.2 &	10.0 &	16.7 \\
\hline
Propulsion &	53.0 & 		5.0 &		55.7 \\
\hline
Harness  &	21.4 &		- &	21.4 \\
\hline
\hline
\textbf{Sum} & \textbf{-} &	\textbf{-} &	 \modif{\textbf{483.8}} \\
\hline
\textbf{Margin} &	\textbf{-} &	\textbf{20.0} &	\modif{\textbf{96.8}} \\
\hline
\textbf{Total} &	\textbf{-} & \textbf{-} &	 \modif{\textbf{580.5}}\\
\hline
\textbf{Maximum} &	\textbf{-} &	\textbf{-} &	\textbf{608.8} \\
\hline
\textbf{Unused} &	\textbf{-} &	\textbf{-} &	\modif{\textbf{28.2}} \\
\hline
\end{tabular}
\begin{list}{}{}
\item[] $^*$ A boom of 3.2~kg has been added to the payload mass estimate (see table \ref{tab:instruments})
\end{list}
\end{table}

\subsection{Power Budget}

The power system consists of three main modules: the primary module, the secondary module and the PCDUs. 

The primary module covers the main power harvesting in order to operate the satellite. The triple junction solar cells with a GaInP2/GaAs/Ge composition (produced, e.g., by Spectrolab), which are designed for space missions, could be used. Each satellite would embark a solar panel area of 2.5~m\textsuperscript{2} featuring an efficiency of 29.5~\%. When the solar panels are exposed to direct sunlight at a distance of 1 AU, the maximum power conversion achievable is calculated to be \modif{905.1}~W with the assumption of surface temperature of 28$^{\circ}$C. The radiation degradation of the solar cells after 7~years of operation time (end of nominal mission) is expected to be approximately 10~\%.

The secondary module consists of the backup power stored in rechargeable lithium ion batteries. For a long lifetime, the batteries are discharged to 50 \% of the full capacity and charged to 90 \% of the full capacity. The total energy capacity of the battery pack is calculated to be 880~Wh with the assumption that the batteries operate at 20$^{\circ}$C.

The distribution of the power is handled by two low power 
PCDUs from ThalesAlenia Space which manage the batteries and the maximum peak power tracking for power harvesting. Each PCDU is able to deliver up to 330~W. 

Table \ref{tab:power} summarises the power budget for each spacecraft of proposed mission OSCAR. The margin for the OBC\&DH including the solid state recorder (SSR) is assumed to be 10~\%. For other sub-systems which still require more modifications a margin of 20~\% is assumed. This applies to TT\&C, EPS or propulsion system. The margins of scientific payload and ADCS result from single margins of the components the sub-systems comprise of based on the standard ESA margin policy. The estimated power for each spacecraft is \modif{855.4}~W including ontop of sub-system margins an additional margin of 20\%, which provides us with an unused power of \modif{49.8}~W per spacecraft.

\begin{table}
\centering
\caption{Power budget. The margins are averaged from the standard ESA margin policy.\label{tab:power}}

\begin{tabular}{|l|c|c|c|}
\hline
\textbf{Subsystem}  & 
\begin{tabular}[c]{@{}c@{}}\textbf{Nominal Power} \\ \textbf{[W]}\end{tabular} 
& 
\begin{tabular}[c]{@{}c@{}}\textbf{Margin} \\ \textbf{[\%]}\end{tabular} 
 & 
 \begin{tabular}[c]{@{}c@{}}\textbf{Sum} \\ \textbf{[W]}\end{tabular} 
\\
\hline
\hline
Payload & \modif{91.8} &		12.3 &		\modif{103.1} \\
\hline
TT\&C  & 200.0 &	20.0 &		240.0 \\
\hline
ADCS  &	205.0&	7.9 &	221.3 \\
\hline
OBC\&DH & 75.0 &	10.0 & 82.5	\\
\hline
Propulsion &	55.0 & 	20.0 &		66.0 \\
\hline
\hline
\textbf{Sum} & \textbf{-} &	\textbf{-} &	 \modif{\textbf{712.8}} \\
\hline
\textbf{Margin} &	\textbf{-} &	\textbf{20.0} &	\modif{\textbf{142.6}} \\
\hline
\textbf{Total} &	\textbf{-} & \textbf{-} & \modif{\textbf{855.4}} \\
\hline
\textbf{Maximum} &	\textbf{-} &	\textbf{-} & \textbf{905.1}\\
\hline
\textbf{Unused} &	\textbf{-} &	\textbf{-} &	\modif{\textbf{49.8}}\\
\hline
\end{tabular}
\end{table}

\subsection{Onboard Computer, Data Handling and Telemetry}
\label{sec:obc}

The data onboard the satellite will be handled by an On Board Computer
(OBC) of the type OSCAR (coincidentally) manufactured by EADS Astrium. The OBC utilises
the LEON3 core and provides up to 40~MIPS at 48~MHz core
frequency. With 256~MB of RAM and 512~MB of exchange memory the
computer meets our requirements. Not only the telemetry data and
command handling but also the execution of the ADCS algorithms and
time synchronization can be performed on the OBC. The processing and
analysis of acquired images is dedicated to a separate image
processing unit. 

The images produced by the scientific instruments will generate a
large amount of data (see section \ref{sec:op_mode_ground_seg}), some
of which will be required to be stored onboard before being downloaded
on Earth. A flight-proven EADS Astrium SSR based on the flash
technology could be utilised to ensure a storage capaticity of
20~Tbits ($\sim$2.5~TB). The power consumption is estimated to be 60~W
and its mass to be 20~kg based on a realistic increase of performances
of SSR in the coming years. These specifications are included in the
corresponding budgets (tables \ref{tab:mass} and \ref{tab:power}) in
the OBC\&DH entry.

The data 
will be downloaded to the ground station (or ground station network)
periodically. Because of the large distance between the spacecraft and the Earth, the telemetry design is particularly critical for the feasibility of OSCAR. In order to provide sufficient downlink budget, X-band
communication is utilised. Two redundant transceivers with the output
power of 200~W each feeding a 1.7~m diameter parabola antenna will
ensure a downlink data rate of 1.4~Mbps if the ESA ESTRACK network is used,
and 260~kbps if smaller 15~m ground station antennas are utilised. The
total daily data budget would then be 2.35~GB and 218~MB respectively. In both
cases the signal margin of 3~dB is maintained in order to guarantee proper
operation. We demonstrate in section~\ref{sec:op_mode_ground_seg} how
such a reasonable telemetry budget is able to meet the scientific
requirements of the mission, and propose alternative budgets depending on the ground stations availability. 

\subsection{Thermal Control System} 

The role of the thermal control subsystem is to maintain all spacecraft and payload components within their required temperature limits during the mission.
Table \ref{tab:thermal} shows the thermal requirements for each component of the spacecraft. The PIM instrument inherits its property from the future PHI instrument that will fly on Solar Orbiter. Its operational and survival temprature ranges are still not exactly known today. Similarly, the exact EUVARI instrument does not exist yet (see section \ref{sec:crit_tech_req}), and its operational temprature ranges can only be speculated. In addition, those two instruments would include their own thermal system. For these reasons, we chose not to consider them in the following preliminary analysis.
    
\begin{table}[htbp]
\centering
\caption{Known thermal requirements for spacecraft and payload components 
\label{tab:thermal}}
\begin{tabular}{|l|c|c|}
\hline
\textbf{Component}  & 
\begin{tabular}[c]{@{}c@{}}\textbf{Operational temperature} \\ \textbf{[$^{\circ}$C]}\end{tabular} 
&
\begin{tabular}[c]{@{}c@{}}\textbf{Survival temperature} \\ \textbf{[$^{\circ}$C]}\end{tabular} \\
\hline
\hline
Batteries$^\mathrm{a}$ 
& -5 to 50  & -15 to 60	
\\
\hline
Reaction wheels$^\mathrm{a}$ 
& -5 to 50 &	-20 to 60 
\\
\hline
Sun sensor$^\mathrm{a}$ 
&	-80 to 95    &	-40 to 90 
\\
\hline
Star trackers$^\mathrm{a}$ 
&	-30 to 60 &	-40 to 70 
\\
\hline
OBC\&DH$^\mathrm{a}$ 
&	-20 to 60 &	-40 to 75 
\\
\hline
Antenna Gimbals$^\mathrm{a}$
&	-40 to 80 &	-50 to 90 
\\
\hline
Antennas$^\mathrm{a}$
&	-100 to 100 &	-120 to 120 
\\
\hline
Solar panels$^\mathrm{a}$
&	-150 to 120 & 		-200 to 130 
\\
\hline
MAG$^\mathrm{b}$
&	-100 to 100 &	-100 to 100 
\\
\hline
SWPM$^\mathrm{c,d}$
&	-25 to 50 &	-30 to 60 
\\
\hline
EPM$^\mathrm{c}$
&	-25 to 50 &	-30 to 60 
\\
\hline
COR$^\mathrm{e}$
& 0 to 40 &	-20 to 55 
\\
\hline
HI$^\mathrm{e}$
&	-20 to 30 &	-60 to 60 
\\
\hline
\end{tabular}

\begin{list}{}{}
\item[] $^\mathrm{a}$ \citet{Wertz:2003vp} $\;$ $^\mathrm{b}$
  \citet{Carr:2007tj} $\;$ $^\mathrm{c}$ \citet{McComas:1998fy} $\;$
  $^\mathrm{d}$ \citet{Sauvaud:2008aa}\\ $^\mathrm{e}$ \citet{Howard:2008fx}
\end{list}
\end{table}

A thermal analysis is generally needed to define an adequate radiator area to accommodate the maximum operational power during the hottest and coldest operational environment, without exceeding the allowed temperatures of  0$^{\circ}$C and 30$^{\circ}$C (see table \ref{tab:thermal}).

As a first approximation it is possible to assume an isothermal and spherical spacecraft with a radius equal to the maximum dimension of the longest subsystem ($\sim$1~m) located at 1~AU from Sun and 0.59~AU from Earth. The satellites will be heated continuously by the direct solar radiation during the whole mission.

The temperature of each spacecraft depends on the balance between its
absorbed, internally diffused and externally radiated thermal power.
The internal power dissipation varies from 99.3~W (safe mode) to
858.2~W (nominal mode). By considering a spacecraft emittance of 0.8
and an absorptivity of the solar radition of 0.6, one finds
equilibrium temperatures of -3.85$^\circ$C and 48.99$^\circ$C,
respectively. A radiator (with an emittance of 0.9) of 3.42~m$^2$
would be required to accomodate the internal dissipation power in
normal operations. This unrealistically large radiator area promotes a
more detailed thermal analysis that could follow two paths. Because
the OSCAR spacecraft will conserve their orientation with respect to
the Sun during the mission, heat shields could be designed on the
Sun-facing part of the spacecraft to lower the internal
temperature. In a perfect-shield case, a radiator of 1.98~m$^2$ would
suffice to maintain the internal temperature in the operational
range. Alternatively, an active internal cooling system could also be
designed to accomodate the internal power dissipation. Finally, the
spacecraft temperature in safe mode is very close to the survival and
operational limits, henceforth a passive thermal control could be
designed to maintain an acceptable temperature in this
case. \modif{This basic thermal analysis shows that 
the design of OSCAR is \textit{a priori} feasible.}

\section{Mission Design}
\label{sec:mission_design}

After the description of the instruments and spacecraft design of OSCAR, we now give insights on the mission design. We detail the possible orbits of the spacecraft and the operational phases in section \ref{sec:orbit_and_phases}. We then propose operational modes for the scientific and forecast data (section \ref{sec:op_mode_ground_seg}). We finally give an overview of a possible ground segment design (section \ref{sec:ground_seg}).

\subsection{Planned Orbit and Operational Phases}
\label{sec:orbit_and_phases}

The OSCAR spacecraft will be inserted into a heliocentric orbit at a distance of 1~AU from the Sun with one spacecraft leading the Earth and other trailing behind with a separation angle of 68$\pm$3$^{\circ}$ (see figure \ref{fig:schematic}). This configuration will allow an optimum observation of the Sun surface to study CMEs and coronal loops, and the optimal acquisition of binocular high resolution images as explained in section \ref{sec:science_req}. Additionally, observation of the CME propagation along the way to the Earth will also be possible.

Five operational phases have been identified for the entire mission: (i) the launch, (ii) a two year spacecraft drift period, (iii) 5 years of  nominal mission time, (iv) a possible mission extension and (v) potential deorbiting. 

\begin{figure}
\centering
\includegraphics[width=0.4\linewidth,angle=-90]{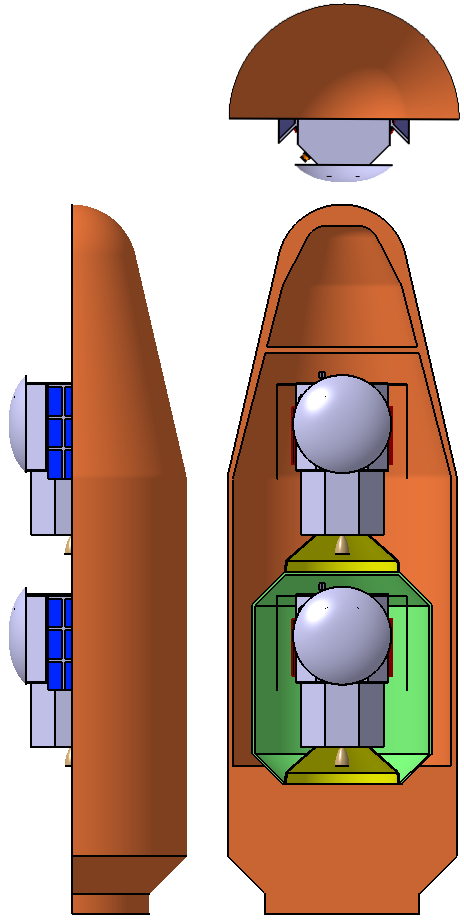}
\caption{The two OSCAR spacecraft stacked in a Soyuz rocket fairing\label{fig:soyuz}}
\end{figure}

The total launch mass, i.e. the sum of dry mass ($\sim$\modif{1161}~kg, see table \ref{tab:mass}), the propellant mass ($\sim$575~kg) and the mass of the payload adapters ($\sim$270~kg), is approximately \modif{2006} kg. A Soyuz rocket, which is capable of delivering 2200 kg to an Earth escape trajectory, has therefore been selected for the mission (the two OSCAR spacecraft also fit in the Soyuz rocket fairing, as shown in figure \ref{fig:soyuz}). After escaping the Earth gravity, the two spacecraft will each perform a 0.47 km/s delta-v manoeuvre in opposite directions i.e. Oscar 1 (leading Earth, see figure \ref{fig:schematic}) will perform a retrograde burn and Oscar 2 (trailing Earth) a prograde burn with respect to their heliocentric orbit. This marks the start of a two years drift phase for both the spacecraft. At the end of this phase each spacecraft will have reached its final target position relative to Earth and will require another 0.47 km/s delta-v manoeuvre to stop the drift in order to maintain the final constellation. The distance between the spacecraft and Earth will be about 0.59 AU.

The launch will take approximately 30 minutes for the spacecraft to reach the Earth escape trajectory. Once they reach this point, their propulsion system will be activated and the drift phase will begin. After about 8 months the angle between the Sun and both spacecraft will reach 22$^{\circ}$ which is the minimum requirement for the spacecraft to start performing observations to achieve part of our mission objectives (section \ref{sec:science_req}). Additionally, at this stage we can start evaluating and if necessary optimise the data handling and propagation of data from the ground stations to various data centres that are mentioned in section \ref{sec:ground_seg}. The scheduled science operation phase is 5 years. However, since we have considered generous margin for the propellant and the power budgets, an extension of the mission duration is optional. After the end of mission both spacecraft will be transferred to a disposal orbit around the Sun with the major semi-axis length of 0.99 AU.

\subsection{Operational Mode and Ground Segment}
\label{sec:op_mode_ground_seg}

\subsubsection{CME Trigger Data}

\begin{figure}
\centering
\includegraphics[width=1\linewidth]{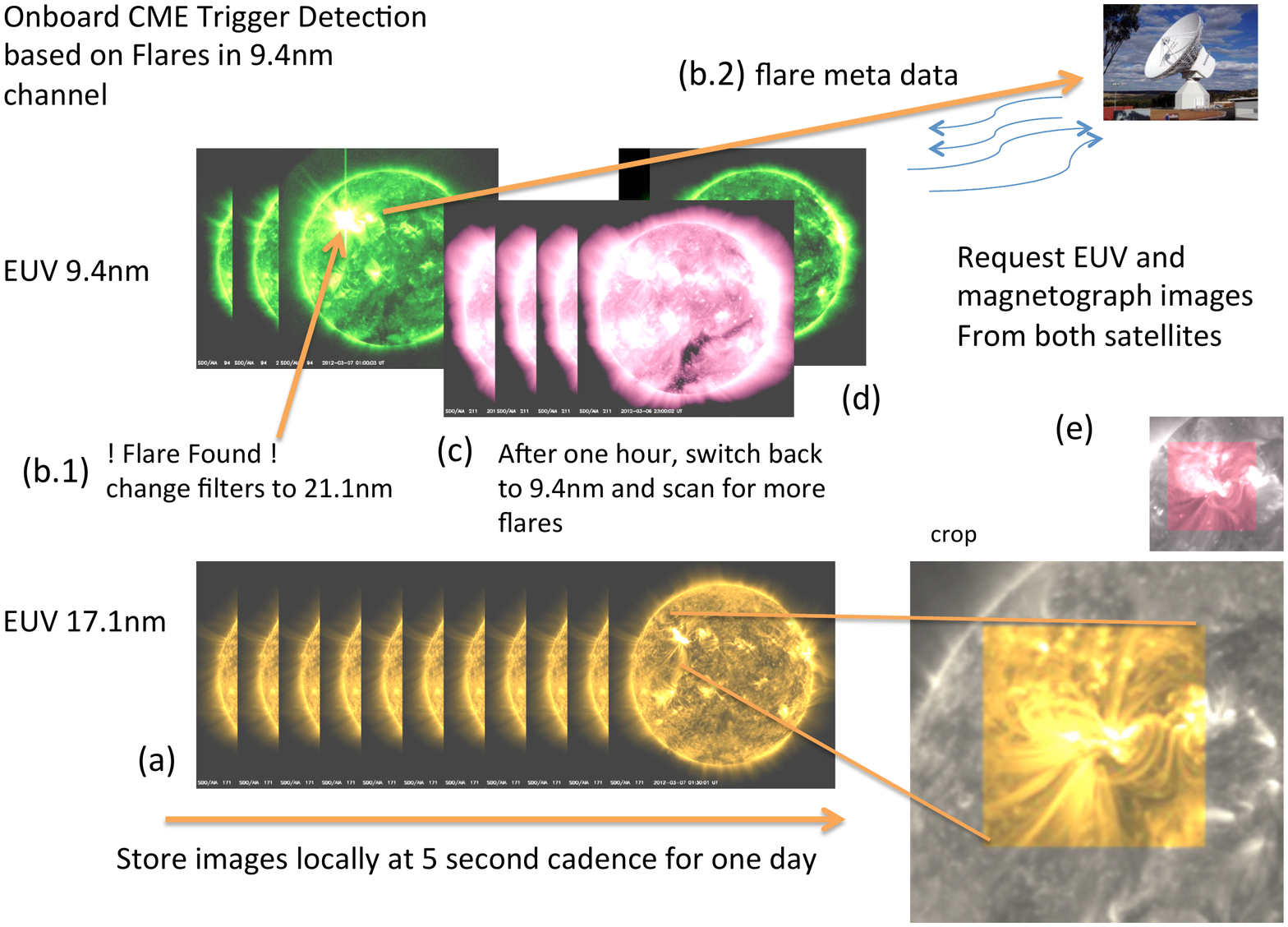}
\caption{OSCAR operational mode for the CME trigger data. See text for details on this particular operational mode.\label{fig:opmode_schematic}}
\end{figure}

Since our primary objective is to study the trigger mechanism(s) of CMEs in high detail, and we can only transmit a limited amount of data to ground stations due to the satellites distance from Earth, onboard autonomy is clearly required. Instead of sending data from all instruments at full cadence and full resolution, our satellites will perform onboard CME trigger event detection using one of the EUV telescopes. We will use a dedicated image processing unit as well as customizable CME trigger event detection software. The following operation mode (summarised in figure \ref{fig:opmode_schematic}) shall enable OSCAR to fulfill its primary objective in spite of the telemetry limitations.

Both EUVARI and PIM telescope shall continously buffer images at full resolution and their fastest cadence (a). The buffering shall be synchronised using ground stations on both satellites taking into account their distance to the Sun. One EUV telescope will record images in 17.1~nm, while the other can switch between the 9.4~nm and the 21.1~nm channel. The trigger detection shall be based on the detection of strong flares in the 9.4~nm channel. Whenever a strong flare is detected (b.1) the satellite will change the 9.4~nm filter to 21.1~nm for the next hour (c). This ensures that we always have 17.1~nm images available, and most of the time right after an event is detected also 21.1~nm images. After one hour the satellite will switch back to the 9.4~nm channel to continue the online flare detection (d).

The output of the event detection is an estimate of the class of the flare, as well as the location. This meta-data will be sent back to a ground station (b.2), where, based on the trigger detection meta-data from both satellites and possibly other sources, is decided which data from both the EUVARI and PIM instruments to request from both satellites (e). It is also possible to request any of the other buffered data, given external trigger detection using third-party data. Furthermore, if simultaneous CMEs were to be triggered, the data could still be retrieved since the EUV telescopes are continuously imaging the full-disk of the Sun.

Other possibilities for online CME trigger event detection can be based on dimmings and EUV waves, instead of on strong flares. Both dimmings and EUV waves are strongly related to the onset of CMEs \citep{Zhukov:2004fa}. An advantage over the flare detector operating on 9.4~nm images would be that both EUV telescopes can continously record in the wavelengths that are best suited for coronal loop imaging (17.1 and 21.1~nm), also for the minutes leading up to CME trigger event. Both EUV waves and dimming detectors as well as flare detectors are currently being developed at the Royal Observatory of Belgium as part of the FP7 project AFFECTS. These can be adapted for near real-time operation on satellites.

Instead of requesting full-resolution data, cropped images will be downloaded for a time period spanning from 10~minutes before the event until 60~minutes after the event was detected. This ensures the total telemetry, given on average 200 M1 or stronger flares per year \citep{Tang:2005vf}, does not exceed 1235 MB per day for each satellite (see the budget including regular and science data in table \ref{tab:telemetry_cme}). \modif{To retrieve this data we make use of the Deep Space Antennas (DSA) of the ESAs ESTRACK network, using one timeslot of 8 hours each day per satellite. The DSA consists of three $35$~m-diameter antennas and is designed to ensure constant availability for spacecraft distant from Earth. It provides the highest telemetry rate available on Earth for our mission. The onboard storage can be viewed as a buffer for the present discussion, containing events of potential interest from which only a portion will be downloaded. The large onboard storage (section \ref{sec:obc}) enables a storage of four days of the whole science data. Depending on the available time-slots of the DSA, more data could be downloaded, other operational modes could be proposed, and calls for allocation time could then be launched. The possibility of using the smaller 15m-diameter antennas of the ESTRACK network (in total six antennas) could also be considered for those imagers as long as this does not affect the downlink time required for the forecasting aspect of the OSCAR mission (see next section). Nevertheless, the 15m-diameter antennas would not support the large amount of the stereoscopic observations by itself and should only be used as complementary antennas.}

\begin{table}[h]
  \begin{center}
\caption{Onboard mass storage and expected telemetry: CME Trigger (for one spacecraft).\label{tab:telemetry_cme}}
    \renewcommand{\arraystretch}{1.2}
\begin{tabular}{|l|r|r|r|r|r|}
 \hline
\textbf{Instrument} & \textbf{Lossless} 
& \begin{tabular}[c]{@{}c@{}}\textbf{Resolution} \\ \textbf{[px]}\end{tabular}  
& \begin{tabular}[c]{@{}c@{}}\textbf{Cadence} \\ \textbf{[s]}\end{tabular}  
& \begin{tabular}[c]{@{}c@{}}\textbf{Onboard storage}\\ \textbf{[MB / day]} \end{tabular} 
& \begin{tabular}[c]{@{}c@{}}\textbf{Telemetry}\\ \textbf{[MB / day]} \end{tabular} \\
\hline\hline
EUVARI  &  y &   4k x 4k &    5 & 552,960 &     \\ 
        &  n &   4k x 4k & 3600 &         & 307 \\ 
        &  y & 800 x 800 &    5 &         & 554 \\ 
\hline
PIM     &  y &   2k x 2k &   45 &  30,720 &     \\ 
        &   &   2k x 2k & 3600 &         & 154  \\ 
        &  y & 400 x 400 &   45 &         &  31 \\ 
        \hline\hline
\textbf{Total} & & & 
& \begin{tabular}[c]{@{}c@{}}\textbf{583,680} \end{tabular} 
& \begin{tabular}[c]{@{}c@{}}  \textbf{1,045} \end{tabular} \\ 
 \hline
\end{tabular}
  \end{center}
\end{table}

\subsubsection{Real-Time Forecasting}

For near real-time forecasting we rely on the availability of 15~meter telescopes that receive data from both satellites every 6 hours. The total telemetry for near real-time forecasting is estimated to be 47~MB every six hours per satellite. This includes telemetry for the coronograph (36~MB), the HI instruments (9~MB), and the in-situ measurements provided by the Particle monitor and Magnetometer (2~MB). This amount of data can be transferred in less than one hour. The 15~m antennas of the ESTRACK network could in principle be used for such telemetry. While this demonstrates the feasibility of OSCAR's forecasting objectives, it could also be considered to develop dedicated infrasctuctures to ensure a \modif{more} reliable forecasting system.

\subsection{Ground Segment}
\label{sec:ground_seg}

\begin{figure}
\centering
\includegraphics[width=0.75\linewidth]{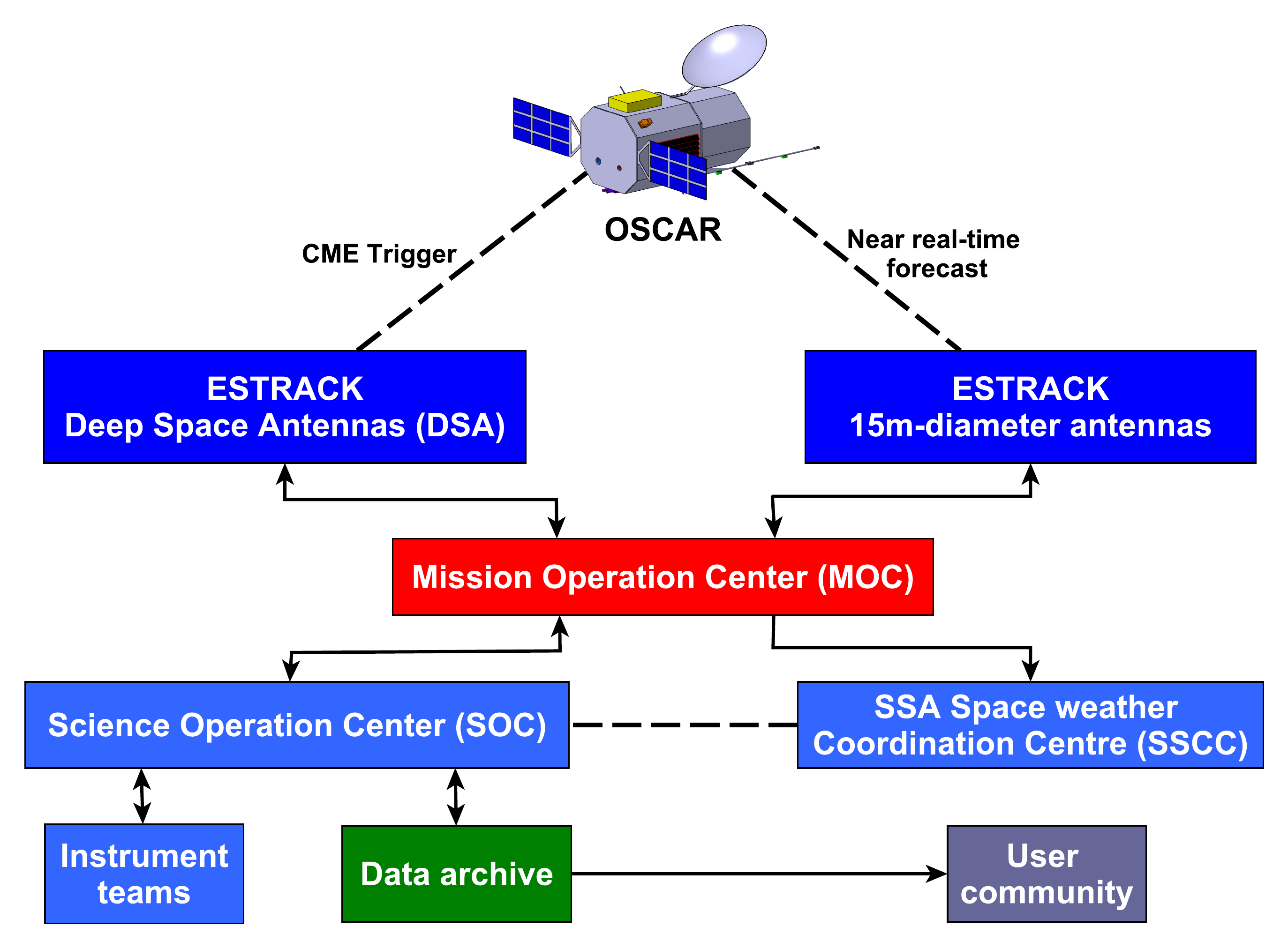}
\caption{OSCAR ground segment design.\label{fig:gd_segment}}
\end{figure}

The operational modes also require a specific design for the mission ground segment. A schematic of our ground segment operation is given in figure \ref{fig:gd_segment}.

\modif{Our design involves the communication with the ESA's ESTRACK network wherein the DSA is used for CME trigger study and the 15~m diameter antennas for the forecasting data.} Our Mission Operation Center (MOC) would provide an interface between the two antenna networks and the science operation and space weather centers. It would also interact with the two satellites for the data downloading requests.

We would of course aggregate a scientific community around our CME trigger study. Specific partners research institutes would be involved in the analysis and use of the data to achieve our first mission objective. 
The forecast data would be directly interfaced with. e.g., the SSA Space weather Coordination Centre (\href{http://swe.ssa.esa.int/web/guest/service-centre}{SSCC}) through our MOC. A constant link between SSCC and the OSCAR SOC would enable a good use of our forecast data. The forecast itself will be either provided by the SOC or by SSCC, depending on the available manpower. Subsequently, SSCC would be in charge of releasing the forecast and alerts obtained from our data. Additionally our forecast data may be combined with other spacecraft data to provide a better insight on the 3D structure and temporal evolution of CMEs and CIRs at 1~AU.

\section{Cost Analysis and Descoping Option}
\label{sec:cost_and_descoping}

We give here rough cost estimates of the OSCAR mission in the evenutallity of a launch at the horizon 2022-2025.
Thanks to the compact design of OSCAR (section \ref{sec:spacecraft_design}), one Soyuz launcher can be used for the two satellites. It has the advantages of a (relatively) low cost and high reliability. A near-equatorial launch location is compatible with the orbit design (section \ref{sec:orbit_and_phases}) and costs 60~M\EUR{} from Kourou. Predictably weight, power, and performance are the main cost drivers. The largest cost in the mission is for the platform (255~M\EUR{}) which represents 57\% of the cost, for which orbit and attitude control and data management represent a large proportion of the platform budget.  The payload cost is also considerable, estimated at 170~M\EUR{} \citep[estimate of 1~M\EUR{}/kg, see][]{Wertz:2003vp}. The development cost of the improved EUV imager is estimated to 5~M\EUR{} by itself. Adding the forecast (100~M\EUR{}) and science (60~M\EUR{}) operations costs, the total cost for the two satellites mission is estimated around 650~M\EUR{}. 

Should it be necessary to reduce the scope of the mission, particularly to reduce the overall cost, the following option has been investigated.
Since both spacecraft have identical instrumentation, a possibility
would be launch only one of them.  For this to retain the objective of
loop reconstruction, this spacecraft would need an orbit relocated at
L5 and would rely on additional image data in the equivalent
wavelengths and cadences from L1 or Earth, using already flying
satellites. If this could be provided by other partners the primary
objective could be fulfilled, while investigating a different portion
of the Sun's surface. However, the matching of suitable data would be
a significant challenge, and data  would likely need to be processed
and interpolated to match in time with the mission data.  In addition,
the data rate would be expected to drop by 64$\%$ compared to the
nominal mission plan due to the additional spacecraft
distance. Although feasible, it clearly appears that this solution
would seriously threaten the mission while saving only a sixth of its
cost.

\section{Conclusions}
\label{sec:conclusions}

We reported a first study for an innovative space weather mission concept, OSCAR. We presented the scientific basis for a twin spacecraft mission, leading and trailing the Earth with a separation angle of 68$^\circ$. OSCAR is designed to answer fundamental questions behind the trigger of CMEs in the lower solar corona, as well as to setup a space weather forecasting system for geoeffective CMEs and CIRs. The advantage of OSCAR resides in the originality of its design that enables to tackle those two goals simultaneously at moderate cost. We furthermore detailed in this work the basic analyses for the feasibility of the OSCAR mission. We put a particular emphasis on showing that, thanks to significant heritage, such a mission requires fairly small instrument developments (the main challenge resides in producing a sufficiently light EUV imager of the lower corona) to lead to important improvements in our scientific understanding of space weather events. In addition, we sketched a full spacecraft design and proposed very simple orbital phases to achieve the required constant angular separation of 68$^{\circ}$ of the two spacecraft (see figure \ref{fig:schematic}). In spite of the large distances involved, the telemetry needed for our mission is accessible with today's terrestrial infrastructures. It must be noted that even though the telemetry requirements may seem demanding, OSCAR would produce the necessary data for very valuable near real-time forecasts of the most dangerous space weather events. In conclusions, the design of the OSCAR mission includes for the very first time real-time predictive capabilities and provides a strong basis for the development of future space weather missions.


\begin{acknowledgements}
  \modif{We thank the referees for constructive reports on our work.
  We acknowledge gratefully the Alpbach Summer School
  organisers and tutors for their invaluable help in the process
  of developing the concept of OSCAR. We acknowledge as well our financial
  supports to participate to the Alpbach Summer School 2013: Centre
  National d'Etudes Spatiales -- CNES (AS
  and TP), Deutsches Zentrum f\"ur Luft- und Raumfahrt -- DLR (NJ,
  AM and MB), UK Space Agency -- UKSA (AL), FFG--Aeronautics and Space
  Agency (PL), FFG-Austrian Research Promotion Agency (BS),
  Finnish Doctoral Programme in Astronomy and Space
  Physics (SH), Royal Observatory of Belgium (EK), Italian Space
  Agency -- ASI (SS), Danish Ministry of
  Higher Education and Science (SM), Enterprise Ireland (LO),
  Ministerio Espa\~{n}ol de
  Econom\'ia y Competitividad (VPB), 
  Norwegian Space Center -- NSC (CS).}
\end{acknowledgements}

\bibliographystyle{swsc}
\bibliography{bib_oscar_SWSC}

\begin{thebibliography}{46}
\providecommand{\natexlab}[1]{#1}
\providecommand{\url}[1]{\texttt{#1}}
\providecommand{\urlprefix}{URL }
\providecommand{\eprint}[2][]{\url{#2}}

\bibitem[{Aschwanden et~al.(2012)Aschwanden, W{\"u}lser, Nitta, and
  Lemen}]{Aschwanden:2012jc}
Aschwanden, M.~J., J.-P. W{\"u}lser, N.~Nitta, and J.~Lemen.
\newblock {Solar Stereoscopy with STEREO/EUVI A and B Spacecraft from Small
  (6∘) to Large (170∘) Spacecraft Separation Angles}.
\newblock \emph{\solphys}, \textbf{281}, 101--119, 2012.
\newblock DOI:
  \href{http://dx.doi.org/10.1007/s11207-012-0092-8}{10.1007/s11207-012-0092-8}.

\bibitem[{Bemporad(2009)}]{Bemporad:2009gf}
Bemporad, A.
\newblock {Stereoscopic Reconstruction from STEREO/EUV Imagers Data of the
  Three-dimensional Shape and Expansion of an Erupting Prominence}.
\newblock \emph{\apj}, \textbf{701}, 298--305, 2009.
\newblock DOI:
  \href{http://dx.doi.org/10.1088/0004-637X/701/1/298}{10.1088/0004-637X/701/1/298}.

\bibitem[{Borovsky and Denton(2006)}]{Borovsky:2006ec}
Borovsky, J.~E., and M.~H. Denton.
\newblock {Differences between CME-driven storms and CIR-driven storms}.
\newblock \emph{\jgr}, \textbf{111}, 2006.
\newblock DOI:
  \href{http://dx.doi.org/10.1029/2005JA011447}{10.1029/2005JA011447}.

\bibitem[{Carr et~al.(2007)Carr, Horbury, Balogh, Baumjohann, Bavassano
  et~al.}]{Carr:2007tj}
Carr, C.~M., T.~S. Horbury, A.~Balogh, W.~Baumjohann, B.~Bavassano, et~al.
\newblock {A Magnetometer For The Solar Orbiter Mission}.
\newblock In Proceedings of The Second Solar Orbiter Workshop, 41, 2007.

\bibitem[{Chen(2011)}]{Chen:2011gq}
Chen, P.~F.
\newblock {Coronal Mass Ejections: Models and Their Observational Basis}.
\newblock \emph{\lrsp}, \textbf{8}, 1, 2011.
\newblock DOI:
  \href{http://dx.doi.org/10.12942/lrsp-2011-1}{10.12942/lrsp-2011-1}.

\bibitem[{Dal~Lago et~al.(2013)Dal~Lago, Gonzalez, De~Lucas, Braga, Vieira,
  Stekel, and Rockenbach}]{DalLago:2013jx}
Dal~Lago, A., W.~D. Gonzalez, A.~De~Lucas, C.~R. Braga, L.~R. Vieira, T.~R.~C.
  Stekel, and M.~Rockenbach.
\newblock {CME dynamics using coronagraph and interplanetary ejecta data}.
\newblock \emph{\adv}, \textbf{51}, 1942--1948, 2013.
\newblock DOI:
  \href{http://dx.doi.org/10.1016/j.asr.2012.11.023}{10.1016/j.asr.2012.11.023}.

\bibitem[{Davies et~al.(2013)Davies, Perry, Trines, Harrison, Lugaz, M{\"o}stl,
  Liu, and Steed}]{Davies:2013iv}
Davies, J.~A., C.~H. Perry, R.~M. G.~M. Trines, R.~A. Harrison, N.~Lugaz,
  C.~M{\"o}stl, Y.~D. Liu, and K.~Steed.
\newblock {Establishing a Stereoscopic Technique for Determining the Kinematic
  Properties of Solar Wind Transients Based on a Generalized Self-Similarly
  Expanding Circular Geometry}.
\newblock \emph{\apj}, \textbf{777}, 167, 2013.
\newblock DOI:
  \href{http://dx.doi.org/10.1088/0004-637X/777/2/167}{10.1088/0004-637X/777/2/167}.

\bibitem[{Davis et~al.(2011)Davis, de~Koning, Davies, Biesecker, Millward
  et~al.}]{Davis:2011cn}
Davis, C.~J., C.~A. de~Koning, J.~A. Davies, D.~Biesecker, G.~Millward, et~al.
\newblock {A comparison of space weather analysis techniques used to predict
  the arrival of the Earth-directed CME and its shockwave launched on 8 April
  2010}.
\newblock \emph{Space Weather}, \textbf{9}, S01,005, 2011.
\newblock DOI:
  \href{http://dx.doi.org/10.1029/2010SW000620}{10.1029/2010SW000620}.

\bibitem[{Fan(2009)}]{Fan:2009kw}
Fan, Y.
\newblock {Magnetic Fields in the Solar Convection Zone}.
\newblock \emph{\lrsp}, \textbf{6}, 4, 2009.
\newblock DOI:
  \href{http://dx.doi.org/10.12942/lrsp-2009-4}{10.12942/lrsp-2009-4}.

\bibitem[{Frazin and Kamalabadi(2005)}]{Frazin:2005aa}
Frazin, R.~A., and F.~Kamalabadi.
\newblock {On the Use of Total Brightness Measurements for Tomography of the
  Solar Corona}.
\newblock \emph{\apj}, \textbf{628}, 1061--1069, 2005.
\newblock DOI: \href{http://dx.doi.org/10.1086/430846}{10.1086/430846}.

\bibitem[{Gandorfer et~al.(2011)Gandorfer, Solanki, Woch, Mart{\'\i}nez~Pillet,
  {\'A}lvarez~Herrero, and Appourchaux}]{Gandorfer:2011dg}
Gandorfer, A., S.~K. Solanki, J.~Woch, V.~Mart{\'\i}nez~Pillet,
  A.~{\'A}lvarez~Herrero, and T.~Appourchaux.
\newblock {The Solar Orbiter Mission and its Polarimetric and Helioseismic
  Imager (SO/PHI)}.
\newblock \emph{Journal of Physics: Conference Series}, \textbf{271}, 2086,
  2011.
\newblock DOI:
  \href{http://dx.doi.org/10.1088/1742-6596/271/1/012086}{10.1088/1742-6596/271/1/012086}.

\bibitem[{Gloeckler et~al.(1998)Gloeckler, Cain, Ipavich, Tums, Bedini
  et~al.}]{Gloeckler:1998fa}
Gloeckler, G., J.~Cain, F.~M. Ipavich, E.~O. Tums, P.~Bedini, et~al.
\newblock {Investigation of the composition of solar and interstellar matter
  using solar wind and pickup ion measurements with SWICS and SWIMS on the ACE
  spacecraft}.
\newblock \emph{\ssr}, \textbf{86}, 497--539, 1998.
\newblock DOI:
  \href{http://dx.doi.org/10.1023/A:1005036131689}{10.1023/A:1005036131689}.

\bibitem[{Gold et~al.(1998)Gold, Krimigis, Hawkins, Haggerty, Lohr, Fiore,
  Armstrong, Holland, and Lanzerotti}]{Gold:1998fc}
Gold, R.~E., S.~M. Krimigis, S.~E. Hawkins, III, D.~K. Haggerty, D.~A. Lohr,
  E.~Fiore, T.~P. Armstrong, G.~Holland, and L.~J. Lanzerotti.
\newblock {Electron, Proton, and Alpha Monitor on the Advanced Composition
  Explorer spacecraft}.
\newblock \emph{\ssr}, \textbf{86}, 541--562, 1998.
\newblock DOI:
  \href{http://dx.doi.org/10.1023/A:1005088115759}{10.1023/A:1005088115759}.

\bibitem[{G{\'o}mez-Herrero et~al.(2011)G{\'o}mez-Herrero, Malandraki, Dresing,
  Kilpua, Heber, Klassen, M{\"u}ller-Mellin, and
  Wimmer-Schweingruber}]{GomezHerrero:2011eo}
G{\'o}mez-Herrero, R., O.~Malandraki, N.~Dresing, E.~Kilpua, B.~Heber,
  A.~Klassen, R.~M{\"u}ller-Mellin, and R.~F. Wimmer-Schweingruber.
\newblock {Spatial and temporal variations of CIRs: Multi-point observations by
  STEREO}.
\newblock \emph{\jastp}, \textbf{73}, 551--565, 2011.
\newblock DOI:
  \href{http://dx.doi.org/10.1016/j.jastp.2010.11.017}{10.1016/j.jastp.2010.11.017}.

\bibitem[{Gopalswamy et~al.(2000)Gopalswamy, Lara, Lepping, Kaiser,
  Berdichevsky, and St~Cyr}]{Gopalswamy:2000ef}
Gopalswamy, N., A.~Lara, R.~P. Lepping, M.~L. Kaiser, D.~Berdichevsky, and
  O.~C. St~Cyr.
\newblock {Interplanetary acceleration of coronal mass ejections}.
\newblock \emph{\grl}, \textbf{27}, 145--148, 2000.
\newblock DOI:
  \href{http://dx.doi.org/10.1029/1999GL003639}{10.1029/1999GL003639}.

\bibitem[{Gosling(1992)}]{Gosling:1992cq}
Gosling, J.~T.
\newblock {In-Situ Observations of Coronal Mass Ejections in Interplanetary
  Space}.
\newblock In Eruptive Solar Flares. Proceedings of Colloquium {\#}133 of the
  International Astronomical Union, 258--267. Springer Berlin Heidelberg,
  Berlin, Heidelberg, 1992.
\newblock ISBN 978-3-540-55246-8.
\newblock DOI:
  \href{http://dx.doi.org/10.1007/3-540-55246-4\_107}{10.1007/3-540-55246-4\_107}.

\bibitem[{Howard et~al.(2008)Howard, Moses, Vourlidas, Newmark, Socker
  et~al.}]{Howard:2008fx}
Howard, R.~A., J.~D. Moses, A.~Vourlidas, J.~S. Newmark, D.~G. Socker, et~al.
\newblock {Sun Earth Connection Coronal and Heliospheric Investigation
  (SECCHI)}.
\newblock \emph{\ssr}, \textbf{136}, 67--115, 2008.
\newblock DOI:
  \href{http://dx.doi.org/10.1007/s11214-008-9341-4}{10.1007/s11214-008-9341-4}.

\bibitem[{Howard and Tappin(2008)}]{Howard:2008hj}
Howard, T.~A., and S.~J. Tappin.
\newblock {Three-Dimensional Reconstruction of Two Solar Coronal Mass Ejections
  Using the STEREO Spacecraft}.
\newblock \emph{\solphys}, \textbf{252}, 373--383, 2008.
\newblock DOI:
  \href{http://dx.doi.org/10.1007/s11207-008-9262-0}{10.1007/s11207-008-9262-0}.

\bibitem[{Kahler et~al.(2005)Kahler, Aurass, Mann, and Klassen}]{Kahler:2005ky}
Kahler, S.~W., H.~Aurass, G.~Mann, and A.~Klassen.
\newblock {The Production of Near-Relativistic Electrons by CME-Driven Shocks}.
\newblock \emph{Coronal and Stellar Mass Ejections}, \textbf{226}, 338--345,
  2005.
\newblock DOI:
  \href{http://dx.doi.org/10.1017/S1743921305000839}{10.1017/S1743921305000839}.

\bibitem[{Liu et~al.(2014)Liu, Luhmann, Kajdi{\v c}, Kilpua, Lugaz
  et~al.}]{Liu2014:aa}
Liu, Y.~D., J.~G. Luhmann, P.~Kajdi{\v c}, E.~K.~J. Kilpua, N.~Lugaz, et~al.
\newblock {Observations of an extreme storm in interplanetary space caused by
  successive coronal mass ejections}.
\newblock \emph{Nature Communications}, \textbf{5}, 3481, 2014.
\newblock DOI: \href{http://dx.doi.org/10.1038/ncomms4481}{10.1038/ncomms4481}.

\bibitem[{Lugaz et~al.(2012)Lugaz, Kintner, M{\"o}stl, Jian, Davis, and
  Farrugia}]{Lugaz:2012gz}
Lugaz, N., P.~Kintner, C.~M{\"o}stl, L.~K. Jian, C.~J. Davis, and C.~J.
  Farrugia.
\newblock {Heliospheric Observations of STEREO-Directed Coronal Mass Ejections
  in 2008 - 2010: Lessons for Future Observations of Earth-Directed CMEs}.
\newblock \emph{\solphys}, \textbf{279}, 497--515, 2012.
\newblock DOI:
  \href{http://dx.doi.org/10.1007/s11207-012-0007-8}{10.1007/s11207-012-0007-8}.

\bibitem[{Manoharan(2006)}]{Manoharan:2006it}
Manoharan, P.~K.
\newblock {Evolution of Coronal Mass Ejections in the Inner Heliosphere: A
  Study Using White-Light and Scintillation Images}.
\newblock \emph{\solphys}, \textbf{235}, 345--368, 2006.
\newblock DOI:
  \href{http://dx.doi.org/10.1007/s11207-006-0100-y}{10.1007/s11207-006-0100-y}.

\bibitem[{Manoharan and Mujiber~Rahman(2011)}]{Manoharan:2011hh}
Manoharan, P.~K., and A.~Mujiber~Rahman.
\newblock {Coronal mass ejections{\textemdash}Propagation time and associated
  internal energy}.
\newblock \emph{\jastp}, \textbf{73}, 671--677, 2011.
\newblock DOI:
  \href{http://dx.doi.org/10.1016/j.jastp.2011.01.017}{10.1016/j.jastp.2011.01.017}.

\bibitem[{Mason et~al.(2009)Mason, Desai, Mall, Korth, Bucik, von Rosenvinge,
  and Simunac}]{Mason:2009cg}
Mason, G.~M., M.~I. Desai, U.~Mall, A.~Korth, R.~Bucik, T.~T. von Rosenvinge,
  and K.~D. Simunac.
\newblock {In situ Observations of CIRs on STEREO, Wind, and ACE During 2007 -
  2008}.
\newblock \emph{\solphys}, \textbf{256}, 393--408, 2009.
\newblock DOI:
  \href{http://dx.doi.org/10.1007/s11207-009-9367-0}{10.1007/s11207-009-9367-0}.

\bibitem[{McComas et~al.(1998)McComas, Bame, Barker, Feldman, Phillips, Riley,
  and Griffee}]{McComas:1998fy}
McComas, D.~J., S.~J. Bame, P.~Barker, W.~C. Feldman, J.~L. Phillips, P.~Riley,
  and J.~W. Griffee.
\newblock {Solar Wind Electron Proton Alpha Monitor (SWEPAM) for the Advanced
  Composition Explorer}.
\newblock \emph{\ssr}, \textbf{86}, 563--612, 1998.
\newblock DOI:
  \href{http://dx.doi.org/10.1023/A:1005040232597}{10.1023/A:1005040232597}.

\bibitem[{M{\"o}stl et~al.(2014)M{\"o}stl, Amla, Hall, Liewer, De~Jong
  et~al.}]{Mostl:2014iv}
M{\"o}stl, C., K.~Amla, J.~R. Hall, P.~C. Liewer, E.~M. De~Jong, et~al.
\newblock {Connecting Speeds, Directions and Arrival Times of 22 Coronal Mass
  Ejections From the Sun to 1 Au}.
\newblock \emph{\apj}, \textbf{787}, 119, 2014.
\newblock DOI:
  \href{http://dx.doi.org/10.1088/0004-637X/787/2/119}{10.1088/0004-637X/787/2/119}.

\bibitem[{Nitta et~al.(2013)Nitta, Aschwanden, Freeland, Lemen, W{\"u}lser, and
  Zarro}]{Nitta:2013iv}
Nitta, N.~V., M.~J. Aschwanden, S.~L. Freeland, J.~R. Lemen, J.~P. W{\"u}lser,
  and D.~M. Zarro.
\newblock {The Association of Solar Flares with Coronal Mass Ejections During
  the Extended Solar Minimum}.
\newblock \emph{\solphys}, \textbf{289}, 1257--1277, 2013.
\newblock DOI:
  \href{http://dx.doi.org/10.1007/s11207-013-0388-3}{10.1007/s11207-013-0388-3}.

\bibitem[{Pulkkinen(2007)}]{Pulkkinen:2007bf}
Pulkkinen, T.
\newblock {Space Weather: Terrestrial Perspective}.
\newblock \emph{\lrsp}, \textbf{4}, 1, 2007.
\newblock DOI:
  \href{http://dx.doi.org/10.12942/lrsp-2007-1}{10.12942/lrsp-2007-1}.

\bibitem[{Richardson(2004)}]{Richardson:2004cx}
Richardson, I.~G.
\newblock {Energetic Particles and Corotating Interaction Regions in the Solar
  Wind}.
\newblock \emph{\ssr}, \textbf{111}, 267--376, 2004.
\newblock DOI:
  \href{http://dx.doi.org/10.1023/B:SPAC.0000032689.52830.3e}{10.1023/B:SPAC.0000032689.52830.3e}.

\bibitem[{Rouillard et~al.(2012)Rouillard, Sheeley, Tylka, Vourlidas, Ng
  et~al.}]{Rouillard:2012ft}
Rouillard, A.~P., N.~R. Sheeley, A.~Tylka, A.~Vourlidas, C.~K. Ng, et~al.
\newblock {The Longitudinal Properties of a Solar Energetic Particle Event
  Investigated Using Modern Solar Imaging}.
\newblock \emph{\apj}, \textbf{752}, 44, 2012.
\newblock DOI:
  \href{http://dx.doi.org/10.1088/0004-637X/752/1/44}{10.1088/0004-637X/752/1/44}.

\bibitem[{Sauvaud et~al.(2008)Sauvaud, Larson, Aoustin, Curtis, M{\'e}dale
  et~al.}]{Sauvaud:2008aa}
Sauvaud, J.~A., D.~Larson, C.~Aoustin, D.~Curtis, J.~L. M{\'e}dale, et~al.
\newblock {The IMPACT Solar Wind Electron Analyzer (SWEA)}.
\newblock \emph{\ssr}, \textbf{136}, 227--239, 2008.
\newblock DOI:
  \href{http://dx.doi.org/10.1007/s11214-007-9174-6}{10.1007/s11214-007-9174-6}.

\bibitem[{Schwenn(2006)}]{Schwenn:2006gg}
Schwenn, R.
\newblock {Space Weather: The Solar Perspective}.
\newblock \emph{\lrsp}, \textbf{3}, 2, 2006.
\newblock DOI:
  \href{http://dx.doi.org/10.12942/lrsp-2006-2}{10.12942/lrsp-2006-2}.

\bibitem[{Shibata and Magara(2011)}]{Shibata:2011kd}
Shibata, K., and T.~Magara.
\newblock {Solar Flares: Magnetohydrodynamic Processes}.
\newblock \emph{\lrsp}, \textbf{8}, 6, 2011.
\newblock DOI:
  \href{http://dx.doi.org/10.12942/lrsp-2011-6}{10.12942/lrsp-2011-6}.

\bibitem[{Simnett et~al.(2002)Simnett, Roelof, and Haggerty}]{Simnett:2002ig}
Simnett, G.~M., E.~C. Roelof, and D.~K. Haggerty.
\newblock {The Acceleration and Release of Near-relativistic Electrons by
  Coronal Mass Ejections}.
\newblock \emph{\apj}, \textbf{579}, 854--862, 2002.
\newblock DOI: \href{http://dx.doi.org/10.1086/342871}{10.1086/342871}.

\bibitem[{Smith et~al.(2004)Smith, Murtagh, and Smithtro}]{Smith:2004kx}
Smith, Z., W.~Murtagh, and C.~Smithtro.
\newblock {Relationship between solar wind low-energy energetic ion
  enhancements and large geomagnetic storms}.
\newblock \emph{\jgr}, \textbf{109}, 1110, 2004.
\newblock DOI:
  \href{http://dx.doi.org/10.1029/2003JA010044}{10.1029/2003JA010044}.

\bibitem[{Stone et~al.(1998)Stone, Frandsen, Mewaldt, Christian, Margolies,
  Ormes, and Snow}]{Stone:1998ea}
Stone, E.~C., A.~M. Frandsen, R.~A. Mewaldt, E.~R. Christian, D.~Margolies,
  J.~F. Ormes, and F.~Snow.
\newblock {The Advanced Composition Explorer}.
\newblock \emph{\ssr}, \textbf{86}, 1--22, 1998.
\newblock DOI:
  \href{http://dx.doi.org/10.1023/A:1005082526237}{10.1023/A:1005082526237}.

\bibitem[{Tang and Le(2005)}]{Tang:2005vf}
Tang, Y.~Q., and G.~M. Le.
\newblock {Statistical Analysis of Soft X-ray Flares during the 23rd Solar
  Cycle}.
\newblock \emph{International Cosmic Ray Conference}, \textbf{1}, 5, 2005.

\bibitem[{Tappin and Howard(2009)}]{Tappin:2009br}
Tappin, S.~J., and T.~A. Howard.
\newblock {Direct Observation of a Corotating Interaction Region by Three
  Spacecraft}.
\newblock \emph{\apj}, \textbf{702}, 862--870, 2009.
\newblock DOI:
  \href{http://dx.doi.org/10.1088/0004-637X/702/2/862}{10.1088/0004-637X/702/2/862}.

\bibitem[{Webb and Howard(2012)}]{Webb:2012er}
Webb, D.~F., and T.~A. Howard.
\newblock {Coronal Mass Ejections: Observations}.
\newblock \emph{\lrsp}, \textbf{9}, 3, 2012.
\newblock DOI:
  \href{http://dx.doi.org/10.12942/lrsp-2012-3}{10.12942/lrsp-2012-3}.

\bibitem[{Wertz and Larson(2003)}]{Wertz:2003vp}
Wertz, J.~R., and W.~J. Larson.
\newblock {Space Mission Analysis and Design}.
\newblock Microcosm ; Kluwer, 3rd illustrated edition edn., 2003.

\bibitem[{Yashiro et~al.(2005)Yashiro, Gopalswamy, Akiyama, Michalek, and
  Howard}]{Yashiro:2005bx}
Yashiro, S., N.~Gopalswamy, S.~Akiyama, G.~Michalek, and R.~A. Howard.
\newblock {Visibility of coronal mass ejections as a function of flare location
  and intensity}.
\newblock \emph{\jgr}, \textbf{110}, 2005.
\newblock DOI:
  \href{http://dx.doi.org/10.1029/2005JA011151}{10.1029/2005JA011151}.

\bibitem[{Yashiro et~al.(2004)Yashiro, Gopalswamy, Michalek, St~Cyr, Plunkett,
  Rich, and Howard}]{Yashiro:2004fo}
Yashiro, S., N.~Gopalswamy, G.~Michalek, O.~C. St~Cyr, S.~P. Plunkett, N.~B.
  Rich, and R.~A. Howard.
\newblock {A catalog of white light coronal mass ejections observed by the SOHO
  spacecraft}.
\newblock \emph{\jgr}, \textbf{109}, 7105, 2004.
\newblock DOI:
  \href{http://dx.doi.org/10.1029/2003JA010282}{10.1029/2003JA010282}.

\bibitem[{Zhang et~al.(2007)Zhang, Richardson, Webb, Gopalswamy, Huttunen
  et~al.}]{Zhang:2007ki}
Zhang, J., I.~G. Richardson, D.~F. Webb, N.~Gopalswamy, E.~Huttunen, et~al.
\newblock {Solar and interplanetary sources of major geomagnetic storms (Dst }.
\newblock \emph{\jgr}, \textbf{112}, 10,102, 2007.
\newblock DOI:
  \href{http://dx.doi.org/10.1029/2007JA012321}{10.1029/2007JA012321}.

\bibitem[{Zhukov and Auchere(2004)}]{Zhukov:2004fa}
Zhukov, A.~N., and F.~Auchere.
\newblock {On the nature of EIT waves, EUV dimmings and their link to CMEs}.
\newblock \emph{\aap}, \textbf{427}, 705--716, 2004.
\newblock DOI:
  \href{http://dx.doi.org/10.1051/0004-6361:20040351}{10.1051/0004-6361:20040351}.

\bibitem[{Zuccarello et~al.(2013)Zuccarello, Balmaceda, Cessateur, Cremades,
  Guglielmino et~al.}]{Zuccarello:2013ia}
Zuccarello, F., L.~Balmaceda, G.~Cessateur, H.~Cremades, S.~L. Guglielmino,
  et~al.
\newblock {Solar activity and its evolution across the corona: recent
  advances}.
\newblock \emph{\jswsc}, \textbf{3}, 18, 2013.
\newblock DOI:
  \href{http://dx.doi.org/10.1051/swsc/2013039}{10.1051/swsc/2013039}.

\bibitem[{Zwickl et~al.(1998)Zwickl, Doggett, Sahm, Barrett, Grubb
  et~al.}]{Zwickl:1998fu}
Zwickl, R.~D., K.~A. Doggett, S.~Sahm, W.~P. Barrett, R.~N. Grubb, et~al.
\newblock {The NOAA Real-Time Solar-Wind (RTSW) System using ACE Data}.
\newblock \emph{\ssr}, \textbf{86}, 633--648, 1998.
\newblock DOI:
  \href{http://dx.doi.org/10.1023/A:1005044300738}{10.1023/A:1005044300738}.

\end{thebibliography}

\end{document}